\documentclass[twocolumn,tighten]{aastex631}

\usepackage{color,comment,multirow,booktabs,upgreek}
\usepackage{multirow,enumitem}
\usepackage[flushleft]{threeparttable}
\usepackage{amsmath}
\usepackage{CJK}
\graphicspath{{./}{images/}}

\newcommand{\Ha}{H_{a}}

\newcommand {\kms}{\ifmmode{\rm km \, s^{-1}}\else{$\rm km \, s^{-1}$}\fi} 
\newcommand {\mo}{{\rm M}_\odot}

\newcommand{\tworef}[2]{\autoref{#1}-\ref{#2}}

\defcitealias{Armillotta+21}{Paper I}
\defcitealias{Armillotta+22}{Paper II}
\defcitealias{Armillotta+24}{Paper III}

\begin{document}
\begin{CJK*}{UTF8}{gbsn}

\title{Energy-Dependent Transport of Cosmic Rays in the Multiphase, Dynamic Interstellar Medium}

\correspondingauthor{Lucia Armillotta}
\email{lucia.armillotta@inaf.it}

\author[0000-0002-5708-1927]{Lucia Armillotta}
\affiliation{INAF Arcetri Astrophysical Observatory, Largo Enrico Fermi 5, Firenze, 50125, Italy}
\affiliation{Department of Astrophysical Sciences, Princeton University, Princeton, NJ 08544, USA}

\author[0000-0002-0509-9113]{Eve C. Ostriker}
\affiliation{Department of Astrophysical Sciences, Princeton University, Princeton, NJ 08544, USA}

\author[0000-0001-8840-2538]{Nora B. Linzer}
\affiliation{Department of Astrophysical Sciences, Princeton University, Princeton, NJ 08544, USA}

\begin{abstract}
We investigate the transport of spectrally resolved cosmic ray (CR) protons with kinetic energies between $1-100$~GeV within the dynamic, multiphase interstellar medium (ISM), using a two-moment CR fluid solver applied to a TIGRESS MHD simulation with conditions similar to the solar neighborhood. Our CR transport prescription incorporates space- and momentum-dependent CR scattering coefficients $\sigma=\kappa^{-1}$, computed from the local balance between streaming-driven Alfv\`{e}n wave growth and damping processes. We find that advection combines with momentum-dependent diffusion to produce a CR distribution function $f(p)\propto~p^{-\gamma}$ with $\gamma\approx4.6$ that agrees with observations, steepened from an injected power law slope $\gamma_\mathrm{inj}=4.3$.
The CR pressure is uniform in the highly diffusive, mostly neutral midplane region, but decreases exponentially in the ionized extraplanar region where scattering is efficient. To interpret these numerical results, we develop a two-zone analytic model that captures and links the two (physically and spatially) distinct regimes of CR transport in the multiphase, dynamic ISM. At low momenta, CR transport is dominated by gas advection, while at high momenta, both advection and diffusion contribute. At high momentum, the analytic prediction for the spectral slope approaches $\gamma=(4/3)\gamma_\mathrm{inj}-1$, and the predicted scaling of grammage with momentum is $X\propto p^{1-\gamma_\mathrm{inj}/3}$, consistent with the simulations. These results support a physical picture in which CRs are confined within the neutral midplane by the surrounding ionized gas, with their escape regulated by both the CR scattering rate in the ionized extraplanar gas and the velocity and Alfv\'{e}n speed of that gas, at effective speed $v_\mathrm{c,eff}\approx(1/2)[\kappa_\parallel~d(v+v_\mathrm{A,i})/dz]^{1/2}$.
\end{abstract}

\vspace*{1cm}

\section{Introduction}
\label{sec:introduction}

Cosmic rays (CRs) are high-energy, charged particles primarily accelerated by shock waves in supernova (SN) remnants \citep[e.g.,][]{Blasi13, Caprioli+23}. Given that their total energy density is comparable to the thermal, turbulent, and magnetic energy densities in the local ISM \citep[e.g.,][]{Boulares&Cox90, Beck01,Grenier+15}, they have the potential to significantly impact gas dynamics in galaxies, including driving galactic winds \citep[e.g.,][]{Zweibel17}. 
Recent theory on dynamical consequences of CRs spans a range from one-dimensional analytic models \citep[e.g.,][]{Mao&Ostriker18, Quataert+21a, Recchia21, Shimoda+22} to magnetohydrodynamic (MHD) simulations of isolated galaxies or cosmological zoom-ins \citep[e.g.,][]{ Hopkins+20, Chan+22, Girichidis+22, Thomas+23} and portions of ISM \citep[e.g.,][]{Girichidis+18, Rathjen+23, Tsung+23, Armillotta+24}. All these studies demonstrate that CR pressure gradients can accelerate galactic outflows, although the efficiency of this process is strongly dependent on the treatment of CR transport \citep[see reviews by][]{Hanasz+21, Ruszkowski+23}. 

Modeling CR propagation on galactic scales is hampered by an incomplete understanding of the microphysical mechanisms coupling CRs to the thermal gas. Being charged particles, CRs stream along magnetic field lines, while scattering off magnetic fluctuations on the scale of their gyroradius. Scattering reduces the effective transport speed, so that for a given CR input rate, stronger scattering enhances the CR confinement in galaxies, resulting in higher CR pressure and greater impacts on the background gas.

The most striking evidence for frequent scattering of CRs in our Galaxy comes from the observed ratios of primary CRs (e.g., H, He, $\alpha$-elements) to secondary CRs (e.g., Li, Be, B), where the latter are produced through spallation of ISM atoms by primary CRs \citep[e.g.,][]{Aloisio&Blasi13, Aguilar+16, Adriani+22}. These ratios are used to infer the CR grammage -- the total ISM mass per unit area traversed by CRs -- which, in turn, provides an estimate of the residence time of CRs in the Galaxy. For CRs at GeV energies, the residence time is estimated to be $\sim 10^7$~yr, suggesting that CRs are confined in the Galaxy for a time much longer than the crossing time for relativistic particles. Additionally, the secondary-to-primary CR ratios, and therefore the CR grammage and residence time, decrease with increasing CR energy, suggesting that high-energy CRs escape the Galaxy more quickly than low-energy CRs. 
This indicates that CRs are not primarily confined by the large-scale wandering of magnetic field lines, as such confinement would result in too little energy dependence of the CR residence time. Instead, this evidence supports the theory that CR confinement is due to scattering off of tiny magnetic fluctuations, which manifests as an energy-dependent diffusive process on large scales \citep[see reviews by][]{Grenier+15, Amato&Blasi18}.

Beyond the secondary-to-primary ratios, additional insights into CR transport in the Galaxy are provided by the CR spectra themselves, which combine the information about the source spectra with the effects of decreasing CR confinement time as energy increases (along with radiative losses for some CR species). Recent modelling of primary CR spectra has suggested that the scattering regime differs at low and high energy \citep[e.g.,][]{Zweibel17, Evoli+18}: sub-GeV and GeV CRs scatter off self-excited Alfv\'{e}n waves (the self-confinement scenario; e.g., \citealt{Kulsrud&Pearce69, Wentzel74}), while ultra-GeV CRs scatter off extrinsic turbulent fluctuations (the extrinsic turbulence scenario; e.g., \citealt{Chandran00, Yan&Lazarian02}). 

Spectral behavior of CRs has a long history of study via traditional phenomenological modeling, making use of code packages such as
\textsc{Galprop} \citep{Strong+98}, \textsc{Dragon} \citep{Evoli+08}, \textsc{Picard} \citep{Kissmann14}, \textsc{Usine} \citep{Maurin+01}, and others \citep[see reviews by][]{Strong+07, Amato&Blasi18, Ruszkowski+23}. These models are highly effective in reproducing observable CR signatures, as they include detailed treatments of CR spectra and non-thermal emission. However, they are limited by their reliance on simplified prescriptions for the thermal gas and magnetic field distributions, as well as for CR transport. The transport process is typically modeled through advection by a galactic wind/fountain and diffusion, both of which are parametrized based on CR observational data, rather than physical properties of the ISM gas. An exciting development in recent years is the incorporation of CR transport within MHD simulations, which are complementary in many ways to traditional phenomenological approaches.

As reviewed e.g.~by \citet{Ruszkowski+23}, most MHD studies of CR-ISM interactions -- where CRs are treated as a relativistic fluid -- have been agnostic about the microphysics of CR scattering and have used a constant scattering (or diffusion) coefficient to parameterize CR propagation. Very recently, however, a few investigations have begun to incorporate new, physically motivated prescriptions of CR scattering in MHD simulations \citep[e.g.,][]{Hopkins2021, Armillotta+21, Sike+24, Thomas+24}. In these studies, the scattering coefficient is computed from the properties of the background gas and the CRs themselves, based on the predictions of the underlying scattering scenario. 

While recent work that uses local quantities in MHD simulations to compute scattering rates represents a significant advance in the physical characterization of CR transport on galactic scales, a limitation that prevents direct connection to observations in most of these studies is the ``single-fluid'' approximation, i.e.\ the assumption that all CRs are protons with kinetic energy of approximately $1$ GeV.
This approach is justified from a dynamical point of view, given that GeV CR protons dominate the CR energy budget. However, allowing for spectrally resolved CRs is critical to test the specific CR transport prescription adopted in the numerical model and expands our overall physical understanding of the problem.    
Comparison with energy-dependent observable CR signatures, including spectra, grammage, and non-thermal emission, provide more stringent tests of the validity of an adopted transport prescription. 

In recent years, CR spectra have been modeled in MHD simulations of ISM either through postprocessing by calculating the steady-state solution \citep[e.g.,][]{Werhahn+21c, Werhahn+21, Werhahn+21b}, or by evolving CRs in time along with the MHD while treating them as a passive particle distribution \citep[e.g.,][]{Yang&Ruszkowski17, Winner+19, Winner+20, Sampson+23}, or in a fully self-consistent manner with CRs and thermal gas dynamically coupled \citep[e.g.,][]{Ogrodnik+21, Girichidis+20, Girichidis+22, Girichidis+24, Hopkins+22a}. To date, the work by \citet{Hopkins+22} is the only study to incorporate physically motivated prescriptions for CR scattering in MHD simulations with spectrally resolved CRs, yet it does not yield agreement between the model predictions and the observed CR spectra.

In this paper, we present results from  simulations of the transport of spectrally resolved CR protons with kinetic energies ranging from 1 to 100 GeV. This study builds on our previous work \citep[see][hereafter \citetalias{Armillotta+21, Armillotta+22, Armillotta+24}]{Armillotta+21, Armillotta+22, Armillotta+24}, in which we implemented a new 
prescription for CR fluid transport based on the self-confinement theory. Our model treats CR transport as a combination of advection by the thermal gas, streaming along the magnetic field at the local ion Alfv\`{e}n speed, and diffusion relative to the wave frame.  The scattering coefficient is computed by balancing wave excitation and damping, leading to a scattering rate that is strongly mediated by the background gas properties. Previously, we incorporated this prescription in the two-moment CR fluid integration module of \citet{Jiang&Oh18} in 
the MHD code \textsc{Athena++} \citep{Stone+20}, and used it to compute the transport of a single GeV-CR fluid within the TIGRESS MHD simulations that model kpc-sized portions of galactic star-forming disks \citep{Kim&Ostriker17, Kim+20}. Crucially, our previous work revealed that the scattering coefficient varies by more than four orders of magnitude depending on the gas properties. This clearly invalidates the common assumption of a spatially-uniform scattering rate, and underscores the need for a detailed ISM representation in CR studies.

For the present investigation, we have extended our former single-fluid transport scheme to track the simultaneous evolution of multiple CRs fluid components, each representing a given energy bin. This novel scheme is used to calculate the transport of 1-100~GeV CR protons within the TIGRESS simulation of the solar neighborhood environment (see \autoref{sec:methods} for a summary of our methods). Our analysis of simulation outcomes and comparison of the inferred CR proton spectrum with that detected on Earth are given in \autoref{sec:sims}. 
Leveraging both the additional constraints and insight from our energy-dependent simulations, we have developed a novel two-zone theoretical model that is able to capture key physical elements and to explain some longstanding observational puzzles.  
\autoref{sec:1Dmodel} presents this two-zone, one-dimensional (1D) steady-state model, and compares its spectral signatures with those of our numerical simulations. 
In \autoref{sec:discuss}, we connect our results to constraints on energy dependence of transport that have been obtained from traditional phenomenological models of CRs, and briefly compare to conclusions obtained from other spectrally resolved CR-MHD simulations.
Finally, in \autoref{sec:summary} we summarize the principal findings of this study.  

\section{Methods}
\label{sec:methods}

Here, we briefly summarize the main feature of the CR transport scheme and its application to the TIGRESS simulation. We refer readers to \citetalias{Armillotta+21}, \citetalias{Armillotta+24}
and a companion paper (Linzer et al.\ 2025, accepted) for further details.

\subsection{Scheme for spectrally resolved-CR transport}
\label{sec:algorithm}

We numerically solve for the transport of multiple CR fluid components, each representing CR protons within a specific range of kinetic energies $E_\mathrm{k}$. Our simulations employ the MHD code package \textit{Athena}++ and evolve each CR fluid independently using the two-moment formalism described in \citet{Jiang&Oh18} and \citetalias{Armillotta+21}. The two-moment equations governing the transport of the $j$-th CR fluid are:
\begin{equation}
\begin{split}
\frac{\partial e_{\mathrm{c},j}}{\partial t} & + \mathbf{\nabla} \cdot \mathbf{F_\mathrm{c,\mathit{j}}} =  - (\mathbf{v} \, + \, \mathbf{v_\mathrm{s,\mathit{j}}} ) \cdot 
\tensor{\mathrm{\sigma}}_\mathrm{tot,\mathit{j}} \cdot   [  \mathbf{F_\mathrm{c,\mathit{j}}} \,+ \\
- \, \mathbf{v} &\cdot (\tensor{{\mathbf{P}}}_\mathrm{c,\mathit{j}} + e_{\mathrm{c},j} \tensor{\mathbf{I}}) ] 
- \Lambda_\mathrm{coll,\mathit{j}} n_\mathrm{H} e_{\mathrm{c},j} + Q_\mathrm{SN,\mathit{j}}
\;,
\end{split}
\label{eq:CRenergy}
\end{equation}
\begin{equation}
\begin{split}
& \frac{1}{v_\mathrm{m}^2} \frac{\partial \mathbf{F_\mathrm{c,\mathit{j}}}}{\partial t} +  \mathbf{\nabla} \cdot \tensor{\mathbf{P}}_\mathrm{c,\mathit{j}} = - \tensor{\mathrm{\sigma}}_\mathrm{tot,\mathit{j}} \cdot [  \mathbf{F_\mathrm{c,\mathit{j}}} +\\
&- \mathbf{v} \cdot (\tensor{{\mathbf{P}}}_\mathrm{c,\mathit{j}} + e_{\mathrm{c},j} \tensor{\mathbf{I}}) ]
- \frac{\Lambda_\mathrm{coll,\mathit{j}} n_\mathrm{H}}{v_\mathrm{p,\mathit{j}}^2} \mathbf{F}_\mathrm{c,\mathit{j}} \;.
\label{eq:CRflux}
\end{split}
\end{equation}
where $e_{\mathrm{c},j}$, $\mathbf{F}_\mathrm{c,\mathit{j}}$, and $\tensor{{\mathbf{P}}}_\mathrm{c,\mathit{j}}$ are, respectively, the energy density, energy flux, and pressure tensor of the $j$-th CR fluid. We assume approximately isotropic CR pressure, so that $\tensor{\mathbf{P}}_\mathrm{c,\mathit{j}} \equiv P_{\mathrm{c},j}\tensor{\mathbf{I}}$, with $P_{\mathrm{c},j} = (\gamma_\mathrm{c} -1) \,e_{\mathrm{c},j} = e_{\mathrm{c},j}/3$, where $\gamma_\mathrm{c} = 4/3$ is the adiabatic index of the relativistic fluid, and $\tensor{\mathbf{I}}$ is the identity tensor. In \autoref{eq:CRflux}, $v_\mathrm{m}$ represents the maximum speed for CR propagation. In principle, $v_\mathrm{m}$ should be equal to the speed of light $c$ for relativistic CRs. However, here, we adopt the ``reduced speed of light'' approach with $v_\mathrm{m} = 10^4 \, \kms$ much larger than any other speed in the simulation  and much lower than $c$ \citep[e.g.,][]{SkinnerOstriker2013, Jiang&Oh18}.

In \autoref{eq:CRflux}, the term $\tensor{\mathrm{\sigma}}_\mathrm{tot,\mathit{j}} \cdot [ \mathbf{F_\mathrm{c,\mathit{j}}} - (4/3) e_{\mathrm{c},j} \mathbf{v}]$ represents the rate of momentum density exchanged between CRs and thermal gas, with $\mathbf{v}$ the gas velocity. In \autoref{eq:CRenergy}, $\mathbf{v} \cdot  \tensor{\mathrm{\sigma}}_\mathrm{tot,\mathit{j}} \cdot   [\mathbf{F_\mathrm{c,\mathit{j}}} - (4/3)e_{\mathrm{c},j} \mathbf{v} ] $ represents the direct CR pressure work done on or by the gas, while $\mathbf{v_\mathrm{s,\mathit{j}}} \cdot  \tensor{\mathrm{\sigma}}_\mathrm{tot,\mathit{j}} \cdot [ \mathbf{F_\mathrm{c,\mathit{j}}} - (4/3) e_{\mathrm{c},j}\mathbf{v} ] $ represents the rate of energy transferred to the gas via damping of gyro-resonant Alfv\'{e}n waves. In the above, $\mathbf{v_\mathrm{s,\mathit{j}}}$ is the streaming velocity of the $j$-th CR fluid, defined to have the same magnitude as the local Alfv\'{e}n speed in the ions $\mathbf{v_{\rm A,i}} \equiv \mathbf{B}/\sqrt{4\pi\rho_\mathrm{i}}$ -- with $\rho_\mathrm{i} $
the ion mass density -- oriented along the local magnetic field $\mathbf{B}$ and pointing down the CR pressure gradient $\nabla P_{\mathrm{c},\mathit{j}}$. 

The tensor $\tensor{\mathbf{\sigma}}_\mathrm{tot,\mathit{j}}$ is the wave-particle interaction coefficient, diagonal in a coordinate system aligned with the local magnetic field. Parallel to the magnetic field, 
\begin{equation}
    \sigma_{\rm tot,\parallel,\mathit{j}}^{-1}= \sigma_\mathrm{\parallel,\mathit{j}}^{-1} + \frac{v_\mathrm{A,i}}{|\hat B \cdot \nabla P_{\mathrm{c},j} |} (P_{\mathrm{c},j} + e_{\mathrm{c},j}) \, ,
\label{eq:sigmatotpar}    
\end{equation}
where the first term is the inverse of the physical scattering coefficient, and the second term is designed such that field-aligned scattering is applied in a frame streaming at the Alfv\'en speed.  
In the directions perpendicular to the magnetic field, $\sigma_{\rm tot,\perp,\mathit{j}}= \sigma_\mathrm{\perp,\mathit{j}}$, representing scattering by unresolved fluctuations in the direction of the mean magnetic field. In this work, we set $\sigma_\mathrm{\perp,\mathit{j}} = 10 \, \sigma_{\parallel,j}$ (see Section 4.3 of \citetalias{Armillotta+21}). We note that with \autoref{eq:sigmatotpar}, the term in square brackets on the right-hand side of \autoref{eq:CRenergy} and \autoref{eq:CRflux} becomes 
\begin{equation}\label{eq:Fterm}
\mathbf{F_\mathrm{c,\mathit{j}}} 
- 4(\mathbf{v} + \mathbf{v}_{\mathrm{A,i}})P_{\mathrm{c},j}. 
\end{equation}

For CRs with $E_\mathrm{k} \lesssim 100$~GeV, we assume scattering is due to Alfv\'en waves excited by the CRs themselves via resonant streaming instability \citep[e.g.,][]{Zweibel17, Evoli+18}. Thus, $\sigma_{\parallel, j}$ is determined by the local balance between Alfv\'en wave excitation and damping mediated by local gas properties, considering both non-linear Landau damping and ion-neutral damping \citep{Kulsrud&Pearce69, Kulsrud05}. The scattering coefficient reduces to
\begin{equation}
\sigma_{\parallel,j} = \frac{\pi}{8} \sqrt{ \frac{\vert \mathbf{{\hat{B}}} \cdot \nabla P_{\mathrm{c},j}\vert}{v_\mathrm{A,i} P_{\mathrm{c},j}} \frac{\Omega_0 c}{ 0.3 v_\mathrm{t,i} v_\mathrm{p,\mathit{j}}^2} \frac{m_\mathrm{p}}{m_\mathrm{i}} \frac{n_{1,j}}{n_\mathrm{i}}}
\label{eq:sigmaNLL}
\end{equation}
in well ionized, low-density gas where nonlinear Landau damping (NLL) dominates, and 
\begin{equation}
\sigma_{\parallel,j} =  \frac {\pi}{8} \, \frac{\vert \mathbf{\hat{B}} \cdot \nabla  P_{\mathrm{c},j}\vert}{v_\mathrm{A,i} P_{\mathrm{c},j}}  \frac{\Omega_0}{ \langle \sigma v \rangle_\mathrm{in}} \, \frac{m_\mathrm{p} (m_\mathrm{n} + m_\mathrm{i})}{ n_\mathrm{n} m_\mathrm{n}  m_\mathrm{i}}  \frac{n_{1,j}}{n_\mathrm{i}} 
\label{eq:sigmaIN}
\end{equation}
in primarily neutral, denser gas where ion-neutral damping (IN) dominates. 

In both of the above, 
$\Omega_0 = e \vert \mathbf{B} \vert / (m_\mathrm{p} c)$ is the cyclotron frequency for $e$ the proton charge and $m_\mathrm{p}$ the proton mass, $m_\mathrm{i}$ is the ion mass, $n_\mathrm{i}$ is the ion number density. The quantity $n_{1,j}$, which has units of number density, depends on the local CR distribution function $f(p)$ in momentum ($p \equiv [(E_\mathrm{k}/c)^2+2E_\mathrm{k} m_\mathrm{p}]^{1/2}$) space as
\begin{equation}
n_{1,j} \equiv 4 \pi p_{1,j} \int_{p_{1,j}}^\infty p f(p) dp\;,
\label{eq:n1}
\end{equation}
where $p_{1,j}=m_p \Omega_0/k$ is the resonant momentum for wavenumber $k$. In the code, $p_{1,j}$ is set equal to the relativistic momentum ${p}_{j}$ associated with the $j$-th CR fluid (see \autoref{sec:tigress}).~In \autoref{Appendix_n1}, we describe how $n_{1,j}$ is computed in the code. 
In \autoref{eq:sigmaNLL}, 
$v_\mathrm{t,i}$ is the ion thermal velocity (which we set equal to the gas sound speed $c_\mathrm{s}$), and $v_\mathrm{p,\mathit{j}} \approx c$ is the CR relativistic velocity, while in \autoref{eq:sigmaIN}, $m_\mathrm{n}$ is the neutral mass, $n_\mathrm{n}$ is the neutral number density, and $\langle \sigma v \rangle_\mathrm{in} \sim 3\times10^{-9}$~cm$^3$~s$^{-1}$ is the rate coefficient for collisions between H and H$^+$ \citep[][Table 2.1]{Draine11}. We emphasize that $\sigma_{\parallel,j}$ depends not only on the properties of the thermal gas, but also on the properties of the CR fluid itself via ${\vert \mathbf{\hat{B}} \cdot \nabla P_{\mathrm{c},j}\vert}/{P_{\mathrm{c},j}}$, $n_{1,j}$, and $v_\mathrm{p,\mathit{j}}$. Hence, because each fluid $j$ represents CRs within a specific range of kinetic energies, the scattering coefficient is both spatially and momentum dependent.

Finally, in \tworef{eq:CRenergy}{eq:CRflux}, the terms $\Lambda_\mathrm{coll,\mathit{j}} n_\mathrm{H} e_{\mathrm{c},j}$ and $\Lambda_\mathrm{coll,\mathit{j}} n_\mathrm{H} \mathbf{F}_\mathrm{c,\mathit{j}}/v_\mathrm{p,\mathit{j}}^2$ represent, respectively, the rates of CR energy density and CR energy flux decrease due to collisional interactions with the ambient gas, where we consider Coulomb, ionization, and hadronic losses (see \autoref{Appendix_loss}). The term $Q_\mathrm{SN,\mathit{j}}$ in \autoref{eq:CRenergy} represents the injected CR energy density per unit time as a consequence of SN events (see \citetalias{Armillotta+21}
and \autoref{sec:tigress}).

We note that the CR transport scheme employed in this study is approximate, as it treats each CR energy component independently, neglecting interactions between different components. A more accurate scheme for spectrally resolved CRs is currently under development (Armillotta \& Ostriker, in prep.), which allows for source terms associated with spectral-dependent adiabatic effects. In \autoref{Appendix_neweq}, we discuss how \tworef{eq:CRenergy}{eq:CRflux} would differ in such a scheme, concluding that while some quantitative results of this work may change, the overall conclusions would remain unaffected.  

\subsection{Application to the TIGRESS MHD simulation}
\label{sec:tigress}

We use the scheme described in \autoref{sec:algorithm} to compute the transport of spectrally resolved CR protons with $1 \leq E_\mathrm{k} \leq 100$~GeV in the TIGRESS MHD simulation modelling a portion of star-forming galactic disk representative of the solar-neighborhood environment (\citealt{Kim&Ostriker17, Kim+20}; see also \citetalias{Armillotta+21, Armillotta+22, Armillotta+24}). To model the CR spectral distribution, we use 5 different CR fluid components. Each component $j$ has an associated momentum $p_{j} = 10^{\,\mathrm{log_{10}} p_\mathrm{0}+j \Delta p/(N-1)}$, with $N=5$, $\Delta p =  \mathrm{log}_\mathrm{10} ( p_\mathrm{N-1} / p_\mathrm{0})$, $p_\mathrm{0} = p (E_\mathrm{k} = 1 \, \mathrm{GeV})$, and $p_\mathrm{N-1} = p(E_\mathrm{k} = 100 \, \mathrm{GeV})$. This corresponds to $p_{j}$ = 2, 5, 13, 36, and 101 GeV/$c$ ($E_\mathrm{k,\mathit{j}}$ = 1, 4, 12, 35, 100 GeV) for $j$ from 0 to 4. The spectral extension of each CR bin is $d\mathrm{ln}p = 0.1$.

Simulations are performed using the same approach as \citetalias{Armillotta+24}: for each selected TIGRESS snapshot, we first compute the transport of CRs in ``post-processing'' mode (i.e., the MHD quantities are frozen in time, while the energy and flux density
of CRs are evolved through space and time until a steady state is reached); then, starting from the postprocessed snapshots, we perform new simulations in which MHD and CRs are evolved together, allowing for the CR backreaction on the gas. These MHD ``relaxation'' simulations are not fully self-consistent in that they do not include self-gravity to follow new star formation, and they do not include injection of energy and momentum in the thermal gas from radiation and supernova feedback. Hence, we run them only for a timescale shorter than the time for the hot gas to be advected out of the domain (a few Myr); see \citetalias{Armillotta+24} for details). 

During the post-processing runs, CR energy is injected near star cluster particles to model effects from SN events. For each CR fluid $j$, the rate of energy injected from each star particle $s$ is $\dot{E}_\mathrm{c, s, \mathit{j}}=\epsilon_\mathrm{c, \mathit{j}} \, E_\mathrm{SN} \,\dot{N}_\mathrm{SN,s}$, where $\epsilon_{\mathrm{c},j}$ is the fraction of SN energy that goes into production of CRs comprising the $j$-th fluid, $E_\mathrm{SN} = 10^{51}$~erg is the energy released by an individual SN event, and $\dot{N}_\mathrm{SN,s}$ is the number of SNe per unit time determined from the \textsc{Starburst99} code based on the age and mass of the star cluster. The fraction $\epsilon_{\mathrm{c},j}$ is computed assuming that (1) 10\% of SN energy goes into production of CRs with $p \geq p_\mathrm{min} = 1$~GeV/c, and (2) the slope of the injected CR distribution function $f_\mathrm{inj}(p)$ is $-4.3$ \citep[see review by][and references therein]{Caprioli+23}. 
The energy density of the $j$-th CR fluid is
\begin{equation}
e_{\mathrm{c},j} \equiv 4 \pi \int_{p_{j_-}} ^{p_{j_+}}  f(p) E(p) p^2 dp  \;,
\label{eq:jth-energy}
\end{equation}
with $p_{j_\pm} = \mathrm{exp}(\mathrm{ln}p_{j}\pm d\mathrm{ln}p/2)$, and $E(p) \equiv E_\mathrm{k}(p) + m_\mathrm{p} c^2$ the total relativistic energy. Hence, the fraction $\epsilon_{\mathrm{c},j}$ can be calculated as 
\begin{equation}
\epsilon_{\mathrm{c},j} = 0.1 \dfrac{   \int_\mathrm{p_\mathit{j_-}} ^{p_{j_+}} f_\mathrm{inj} E p^2 dp} {\int_\mathrm{p_\mathrm{min}} ^\infty f_\mathrm{inj} E p^2 dp} = 0.1 \dfrac {\int_\mathrm{p_\mathit{j_-}} ^{p_{j_+}} E p^{-2.3} dp} {\int_\mathrm{p_\mathrm{min}} ^\infty E p^{-2.3} dp}\;.
\label{eq:injrate}
\end{equation}
During the MHD relaxation runs, CR energy injection is turned off for consistency with the absence of thermal energy and momentum injection (see above).

\section{Simulation outcomes}
\label{sec:sims}

\begin{figure*}
\centering
\includegraphics[width=\textwidth]{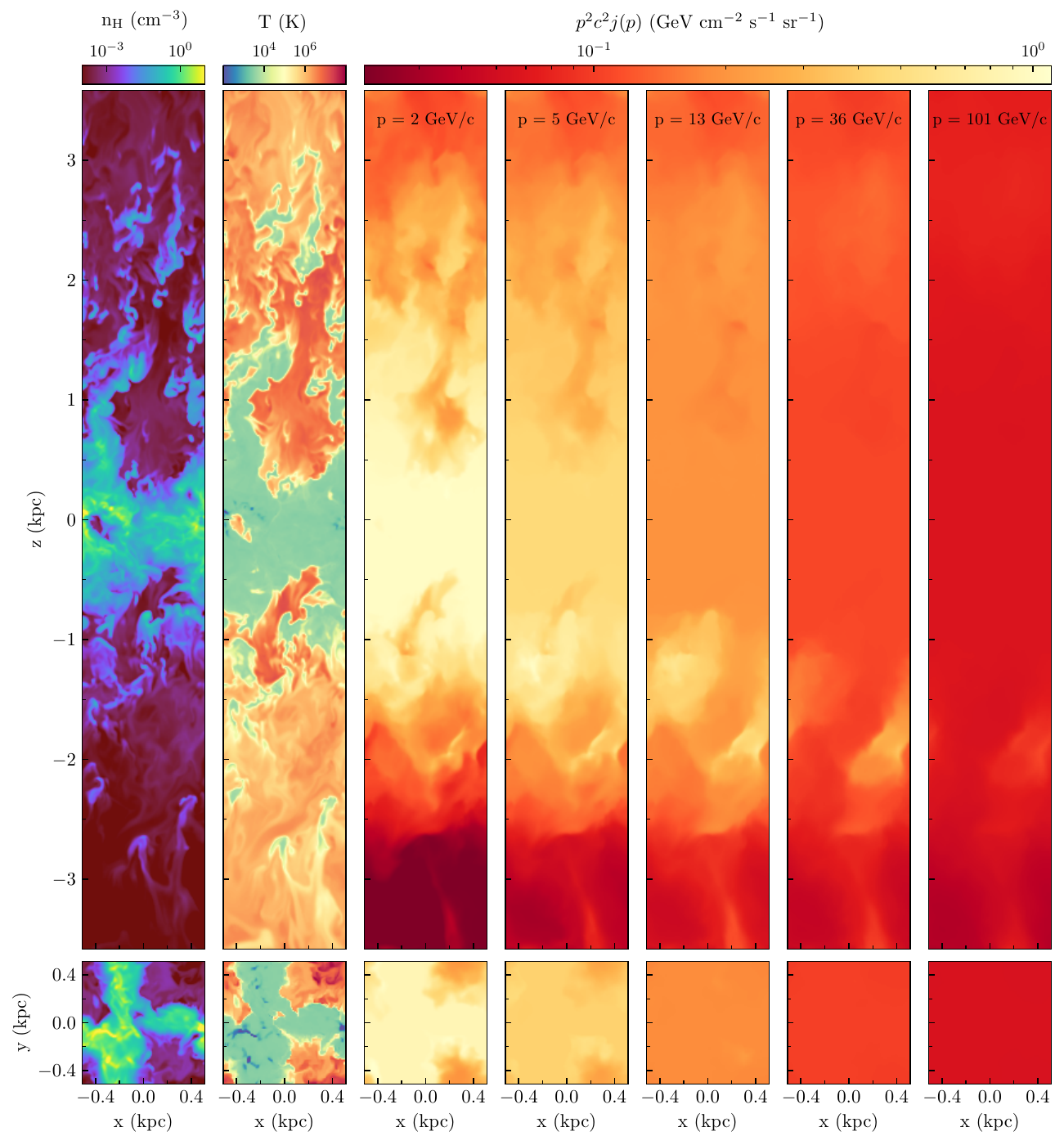}
\caption{Sample snapshot taken at the end the MHD relaxation run. The upper (lower) row of panels shows $x$-$z$ ($x$-$y$) slices through the center of the simulation box, where $x$, $y$, and $z$ are the local radial, azimuthal, and vertical directions.  
From left to right, columns show hydrogen number density $n_\mathrm{H}$, gas temperature $T$, and CR spectral fluxes $j$ at different momenta $p$ multiplied by the square of $pc$.}
\label{fig:FBvel}
\end{figure*} 

In this section, we analyse the spatial and spectral distribution of CRs as computed from the simulations. Hereafter, we shall categorize different CR bins based on their momentum rather than their kinetic energy, consistent with the common approach in observational studies that describe CR spectral properties in terms of momentum (or rigidity, i.e., momentum in units $e/c$). The CR spectral flux $j(p)$ is related to the CR distribution function $f(p)$ as $j(p) = p^2 f(p)$. Assuming that $f$, $E$, $p$ are constant within the integral of \autoref{eq:jth-energy} -- a reasonable approximation given that $d\mathrm{ln}p$ is sufficiently small -- the CR spectral flux associated with the $j$-th CR bin is computed as $j_{j} = e_{\mathrm{c},j}/(4 \pi E_{j} p_{j} d\mathrm{ln}p)$. 

The leftmost two panels of \autoref{fig:FBvel} display the distribution on the grid of gas hydrogen density $n_\mathrm{H}$ and temperature $T$ in one sample TIGRESS snapshot at the end of the MHD relaxation step. From left to right, the remaining five panels show the distributions of CR energy spectrum $p^2 c^2 j(p)$ at momentum values $p = 2$, 5, 13, 36, and 101 GeV$/c$, respectively. The main evidence of \autoref{fig:FBvel} is that the spatial distribution of CRs becomes more and more uniform with increasing CR momentum. As will be discussed later in this Section, this is due to the increasing efficiency of CR diffusion.

\begin{figure*}
\centering
\includegraphics[width=\textwidth]{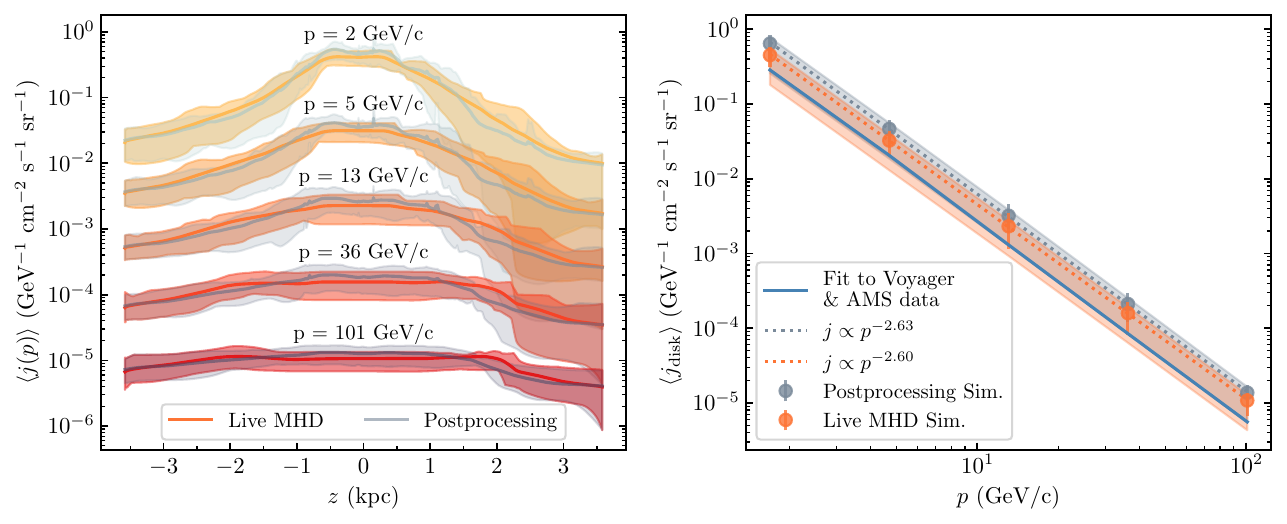}
\caption{Distribution of CR spectral flux $j$ in physical and momentum space. The left panel shows the horizontally averaged vertical profiles of $j$ for the five different momenta investigated in the paper. The right panel shows the average CR spectrum evaluated in the disk region ($\vert z \vert < 500$~pc). The results in gray indicate averages from postprocessing simulations, while the results in red/orange indicate averages from the MHD relaxation simulations at $t=2$~Myr. In the left panel, the shaded areas cover the 16th and 84th percentiles of fluctuations. In the right panel, the points indicate the simulation outcomes (with error bars showing the 16th-84th percentiles), while the dotted lines represent power-law fits. The blue line shows a fit to data for direct CR detections in the solar system \citep{Padovani+18}. The shaded areas cover variations of the simulated spectrum as would apply if the SFR surface density $\Sigma_\mathrm{SFR}$ in the solar neighborhood had increased by a factor $3$ over the last 100 Myr \citep{Zari+23}, to reach a present-day value equal to that in our simulations (with $j \propto \Sigma_\mathrm{SFR}$).}
\label{fig:spectrum}
\end{figure*}

The left panel of \autoref{fig:spectrum} shows the horizontally averaged profiles of CR flux $j(p)$ for different momenta, with the average computed over 8 snapshot outputs. The data in gray indicate averages from the postprocessing simulations, while the data in red/orange indicate averages from the MHD relaxation simulations at $t=2$~Myr. The profiles before and after turning on MHD are overall similar, with the latter being slightly smoother than the former. In the postprocessing runs, the
prevailing orientations of the velocity and magnetic field lines
confine CRs within the warm/cold dense gas. However, once MHD is ``live'', the backreaction
of the CR pressure on the gas rearranges the velocity and
magnetic field topology, enabling CRs to propagate out of the dense gas. This results in a more uniform CR distribution across different thermal phases of the gas (see \citetalias{Armillotta+24}). For every CR momentum, $j$ decreases with $\vert z \vert$. Moreover, as noted above, its vertical profile becomes smoother with increasing $p$.

The distribution of CRs in momentum space is displayed in the right panel of \autoref{fig:spectrum}, showing the mean value of $j$ as a function of $p$ measured within the disk region ($\vert z \vert < 500$~pc). In the plot, the dotted lines indicate the fits to the simulation outcomes (shown as  points with error bars). We find $j \propto p^{-2.63}$ ($f \propto p^{-4.63}$) and $j \propto p^{-2.6}$ ($f \propto p^{-4.6}$) for postprocessing and MHD relaxation simulations, respectively. The inferred spectral slope is in excellent agreement with the fit to the CR proton spectrum measured in the solar system, highlighted by a solid blue line \citep[from][]{Padovani+18}. This fit is a broken power law that peaks at $E_\mathrm{k} = 650$~MeV, with a high-energy slope of $-2.7$.
Notably, the ambient CR spectrum is predicted to be steeper than the injection spectrum ($j_\mathrm{inj} \propto p^2 f_\mathrm{inj} \propto p^{-2.3}$).

The normalization of the CR spectrum computed from our simulations is a factor of $\sim 1.5-2$ higher than the observed one. We note that the total CR energy density is nearly linearly proportional to the SFR surface density $\Sigma_\mathrm{SFR}$ (see \citetalias{Armillotta+22}). The average value of $\Sigma_\mathrm{SFR}$ in the TIGRESS snapshots analysed in this work is $\sim 5\times 10^{-3}\,\mo$~kpc$^{-2}$~yr$^{-1}$, a value consistent with the present-day $\Sigma_\mathrm{SFR}$ in the solar neighborhood, but higher than some empirical estimates of the mean $\Sigma_\mathrm{SFR}$ over the last $50-100$ Myr \citep[see Figure 5 in][]{Zari+23}. Considering that the lifetimes of massive stars evolving to SNe is $\approx 3-40$ Myr and that the typical CR escape time from the Galaxy is $\approx 30$~Myr at GeV energies \citep[e.g.,][]{Ruszkowski+23}, it is reasonable to assert that CRs currently detected in the solar system were produced by SN events from stars formed over the last $50-100$ Myr. In the plot, the shaded areas shows how the normalization of the predicted spectrum would vary based on a factor-of-3 variation in $\Sigma_\mathrm{SFR}$ over this period. The observed spectrum lies perfectly within the shaded area.

\begin{figure*}
\centering
\includegraphics[width=\textwidth]{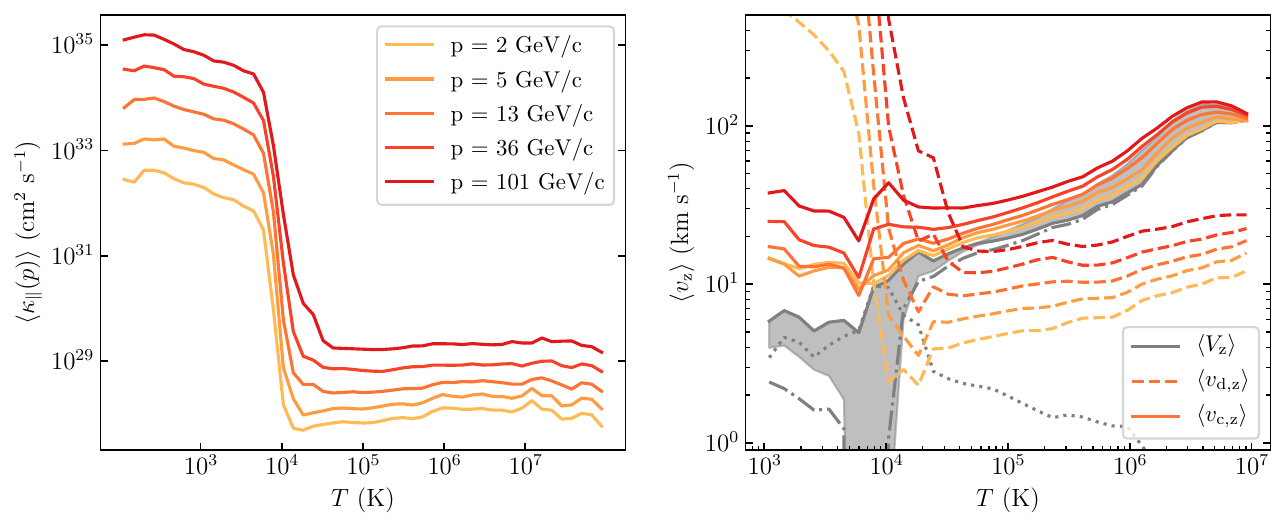}
\caption{CR transport properties at different momenta $p$. The left panel shows the medians of the diffusion coefficients $\kappa_\parallel$ as a function of temperature $T$. In the right plot, the solid and dashed lines indicate the CR pressure-weighted mean vertical components of the effective CR propagation speeds $v_\mathrm{c,z} \equiv F_\mathrm{c,z}/(4P_\mathrm{c})$ and diffusion speeds $v_\mathrm{d,z} \equiv F_\mathrm{d,z}/(4P_\mathrm{c})$, respectively. The gray dotted, dash-dotted, and solid lines represent the CR pressure-weighted mean vertical components of the streaming velocity $v_\mathrm{s,z}$, advection velocity $v_\mathrm{z}$, and MHD velocity $V_\mathrm{z} \equiv v_\mathrm{z} + v_\mathrm{s,z}$ for CRs with momentum $p=2$~GeV/c, while the gray shaded area covers variations of $V_\mathrm{z}$ at different momenta. In both panels, different colors represent different CR momentum bins.}
\label{fig:sigma_vel}
\end{figure*}

The steepening of the CR spectrum compared to the injected spectrum is due to more rapid escape from the galaxy of high energy CRs. 
\autoref{fig:sigma_vel}
explains this effect based on differences in local CR transport at different energies, also showing the dependence on gas thermal properties. The left panel shows the CR diffusion coefficient $\kappa_\parallel \equiv 1/\sigma_\parallel$ (see \autoref{eq:sigmaNLL} and \autoref{eq:sigmaIN}) as a function of gas temperature $T$ at different momentum values. The dependence of $\kappa_\parallel$ on $T$ is qualitatively similar for every $p$: at low temperatures ($T < 10^4$~K), where gas is mainly neutral and IN damping dominates, $\kappa_\parallel$ is large and slightly decreases with $T$; near $T\approx10^4$~K, where gas becomes mostly ionized, $\kappa_\parallel$ plummets by more than four orders of magnitude; at high temperatures, where NLL damping dominates, $\kappa_\parallel$ is almost independent of $T$. For any given $T$, the value of $\kappa_\parallel$ is higher for larger $p$, meaning that CRs are more diffusive at higher energy.  

The momentum dependence of $\kappa_\parallel$ (or $\sigma_\parallel$) is mostly\footnote{The dependencies on the CR velocity $v_\mathrm{p}$ ($\approx c$ for $E_\mathrm{k} > 1$~GeV) and the local CR scale height ${P_\mathrm{c}}/{\vert \mathbf{\hat{B}} \cdot \nabla  P_\mathrm{c}\vert}$ on $p$ are weaker compared to $n_1$.} due to the factor $n_\mathrm{1}$, with \autoref{eq:n1} giving $n_\mathrm{1} \propto p_\mathrm{1}^{3-\gamma}$ for a generic distribution function $f\propto p^{-\gamma}$ (see \autoref{eq:n1_appendix}). 
Thus, we have 
\begin{equation}\label{eq:kappa_scaling}
\kappa_\parallel \propto \begin{cases}
			 p^{(\gamma-3)/2}&  \text{NLL regime}\\
             p^{\gamma-3}\, & \text{IN regime}
		 \end{cases}
\end{equation}
for the scaling of the parallel diffusion coefficient with CR momentum (or energy). In particular, $\kappa_\parallel$ increases with $p$ if $\gamma > 3$, a condition that is always satisfied in our simulations, where the input slope is $\gamma = 4.3$, and the evolved slope is $\gamma \sim 4.6$ (see right panel of \autoref{fig:spectrum}).

The right panel of \autoref{fig:sigma_vel} displays the CR pressure-weighted mean profiles of the vertical components of the MHD velocity $\mathbf{V}$, diffusive CR velocity $\mathbf{v}_\mathrm{d}$, and effective CR velocity $\mathbf{v}_\mathrm{c}$ as a function of $T$ for different $p$. For each CR fluid $j$, the MHD velocity is defined as the sum of gas advection velocity and CR streaming velocity: $\mathbf{V}_{j}  \equiv \mathbf{v} + \mathbf{v}_\mathrm{s,\mathit{j}}$, while the effective CR velocity is defined as the ratio of CR energy flux and energy density: $\mathbf{v}_\mathrm{c,\mathit{j}} \equiv 3/4 \, \mathbf{F}_\mathrm{c,\mathit{j}}/e_{\mathrm{c},j}$. The diffusive velocity is defined such that in steady-state and for negligible collisional losses $\mathbf{v}_\mathrm{c,\mathit{j}}$ reduces to $\mathbf{V}_\mathit{j} + \mathbf{v}_\mathrm{d,\mathit{j}}$. Specifically, $\mathbf{v}_\mathrm{d,\mathit{j}} \equiv 3/4 \, \mathbf{F}_\mathrm{d,\mathit{j}}/e_{\mathrm{c},\mathit{j}}$, where $\mathbf{F}_\mathrm{d,\mathit{j}} \equiv - \tensor{\sigma}_\mathit{j}^{-1} \cdot \nabla P_{\mathrm{c},j}$ is obtained by applying the above assumptions to \autoref{eq:CRflux} and subtracting the advective and streaming fluxes from the total CR flux. 

\autoref{fig:sigma_vel} shows that CRs and thermal gas are well coupled ($v_\mathrm{c,z} \sim V_\mathrm{z}$) in regions of  well-ionized warm-hot ($T>10^{5}$~K) gas, where CR scattering is highly effective (low values of $\kappa_\parallel$). In this regime, advection is the dominant CR transport mechanism. Streaming 
speeds exceed advection for 
$T\lesssim 10^4$~K, but for much of this regime, CRs diffuse very rapidly due to strong IN damping in neutral gas. Thus, $v_\mathrm{d,z} \gg V_\mathrm{z}$ in the neutral gas at $T<10^4$~K. 

Although the diffusion speed is very high ($v_\mathrm{d,z} \gg 10^2 \, \kms$) in  poorly ionized regions, the effective CR transport speed remains significantly lower and comparable to the transport speed at $T\sim10^{4-5}$~K. This is because the propagation of CRs out of the poorly ionized, warm-cold gas in the midplane region (most of the ISM by mass) is limited by the low transport speed in the high-ionization, low-density galactic fountain gas that surrounds the midplane layer. As we shall explain in detail in \autoref{sec:1Dmodel}, for the galactic fountain gas all three velocities are significant, with diffusion becoming increasingly more important at higher $p$.
 
\section{One-dimensional Model}
\label{sec:1Dmodel}

\subsection{Theoretical Formulation}

In this section, we develop a 1D steady-state model for the CR pressure and flux in different spectral bins. In deriving the model, we consider spatial variations and net flux along the $z$-direction only, which is 
a good approximation in the extra-planar region (see \citetalias{Armillotta+21}); this is essentially equivalent to a temporal and horizontal average of \tworef{eq:CRenergy}{eq:CRflux}. 

The 1D steady-state versions of \autoref{eq:CRenergy} and \autoref{eq:CRflux} in a given spectral bin $j$ (where we suppress this index for cleaner notation here) are
\begin{equation}
 \dfrac{ d F_\mathrm{c}}{dz}   =  {V}  \dfrac{ d {P_\mathrm{c}}}{dz} + Q_\mathrm{SN} \;,
\label{eq:CRenergy1D}
\end{equation}
\begin{equation}
\dfrac{ d P_\mathrm{c}}{dz} = - \sigma_\mathrm{\parallel} \left( F_\mathrm{c} - 4 V P_\mathrm{c} \right)\;.
\label{eq:CRflux1D}
\end{equation}
Here $V$ is the sum of the vertical advection and streaming velocity, and $F_\mathrm{c}$ is the vertical component of the flux. 
To obtain \autoref{eq:CRflux1D}, we have used the expression in \autoref{eq:Fterm} for the flux in the wave frame in the scattering term.
In both equations, we have neglected the collisional loss terms $\Lambda_\mathrm{coll} n_\mathrm{H} {e}_\mathrm{c} $ and $\Lambda_\mathrm{coll} n_\mathrm{H} {F}_\mathrm{c}/{v_\mathrm{p}^2}$ because, in the solar neighborhood environment, these losses are negligible compared to the other source/sink terms for CR protons (see Table~2 in \citetalias{Armillotta+22}). We note that both $Q_\mathrm{SN}$ and $\sigma_\parallel$ depend on the CR momentum $p$. In principle, even the MHD velocity $V$ depends on $p$, since the direction of the streaming velocity is determined by the direction of the CR pressure gradient. However, for simplicity, we assume that $V$ is independent of $p$, which is a valid approximation in the regime where transport is primarly regulated by the MHD velocity (see right panel of \autoref{fig:sigma_vel}). 

From \autoref{eq:CRflux1D}, we can derive the steady-state expression for the flux: $F_\mathrm{c}  = - 
\kappa_\parallel d P_\mathrm{c}/dz +  4 V P_\mathrm{c} $, where $\kappa_\parallel = 1/\sigma_\parallel$. We define an effective CR vertical velocity, $v_\mathrm{c,eff}$, and an effective CR diffusion coefficient, $\kappa_\mathrm{eff}$, as

\begin{equation}
v_\mathrm{c,eff} \equiv \dfrac{F_\mathrm{c}}{4P_\mathrm{c}} = V -\dfrac{\kappa_\parallel}{4}\dfrac{d\ln P_\mathrm{c}}{dz}  \;,
\label{eq:veff}
\end{equation}
\begin{equation}
\kappa_\mathrm{eff} \equiv - \dfrac{F_\mathrm{c}}{dP_\mathrm{c}/dz} ={\kappa_\parallel} -\dfrac{4V}{d\ln P_\mathrm{c}/dz} \;.
\label{eq:kappaeff}
\end{equation} 
That is, the effective speed and effective diffusion coefficient allow for all forms of CR transport. 

From the left panel of \autoref{fig:spectrum}, the pressure profiles transition from a flat pattern at lower altitudes, where high CR diffusion in the volume-filling cold/warm mostly-neutral gas smooths out CR inhomogeneities, to an exponential pattern at higher altitudes, where gas is mostly ionized and the CR scattering rate is high. The specific location along the $z$-axis where this transition occurs varies across different spectral bins. Motivated by these results, we adopt as an \textit{Ansatz} for each spectral bin a piecewise function of the following form:
\begin{equation}
P_\mathrm{c}(\vert z \vert) = 
\begin{cases} 
    P_\mathrm{0} \;\;\;\;\;\;\;\;\;\;\;\;\;\;\;\;\;\;\;\;\;\;\;\;\;\;\;\;\;\;\; \mathrm{for} \; \vert z \vert \leq z_\mathrm{t}\\
    P_\mathrm{0} \, \mathrm{exp} \left ({-\dfrac{\vert z \vert - z_\mathrm{t}}{H_\mathrm{c}}} \right) \;\;\;\, \mathrm{for} \; \vert z \vert > z_\mathrm{t}
\end{cases}
\label{eq:piecewisePc}
\end{equation}
with $P_\mathrm{0}$, $z_\mathrm{t}$, and $H_\mathrm{c}$ varying with the CR momentum $p$. 
In the extraplanar region $|z|>z_\mathrm{t}$, the scale height $H_\mathrm{c}$ will differ in each CR energy bin, and the corresponding effective transport velocity and diffusion coefficient are given by 
\begin{equation}\label{eq:veff1}
v_\mathrm{c,eff} = V + \frac{\kappa_\parallel }{4 H_\mathrm{c}} 
\end{equation}
\begin{equation}\label{eq:keff1}
\kappa_\mathrm{eff}= \kappa_\parallel + 4  H_\mathrm{c} V= 4 H_\mathrm{c} v_\mathrm{c,eff} \,. 
\end{equation}

We proceed to seek an analytic solution for $P_\mathrm{0}$ and $H_\mathrm{c}$ by solving \tworef{eq:CRenergy1D}{eq:CRflux1D}. Hereafter, we will refer to the region at $\vert z \vert \leq z_\mathrm{t}$ as ``zone~0'', and to the region at $\vert z \vert > z_\mathrm{t}$ as ``zone~1''. In zone~0, which consists primarily of warm and cold neutral gas, the scattering coefficient is small (see left panel of \autoref{fig:sigma_vel}), implying that $d P_\mathrm{c}/{dz} \approx 0$ and $d F_\mathrm{c}/{dz} = Q_\mathrm{SN}$. Integrating the latter equation from $z = 0$ to $z = z_\mathrm{t}$, and for $F_\mathrm{c} \approx 0$ at the midplane, we obtain $F_\mathrm{c} (z_\mathrm{t}) = F_\mathrm{in} = \dot{E}_\mathrm{in}/(2A)$, where $F_\mathrm{in}$ is the input (or injected) flux, $\dot{E}_\mathrm{in}$ is the sum of the rates of energy injected from each star particle (see \autoref{sec:tigress}), and $A$ is the area of the simulation box.
From \autoref{eq:jth-energy} and \autoref{eq:injrate}, $F_\mathrm{in} \propto p^{4-\gamma_\mathrm{inj}}$, with $\gamma_\mathrm{inj}$ the slope of the input CR distribution function, equal to $4.3$ in our simulations.

In zone~1, away from the midplane, the extraplanar gas is well ionized and the scattering coefficient becomes much larger, while there are no CR sources ($Q_\mathrm{SN}$ = 0). Solving for the flux from \autoref{eq:CRflux1D}, and substituting the solution in \autoref{eq:CRenergy1D}, we obtain the following second-order differential equation: 
\begin{equation}
\kappa_\mathrm{\parallel}\dfrac{d^2}{dz^2}P_\mathrm{c} - 3 V \dfrac{d}{dz}P_\mathrm{c} - 4 P_\mathrm{c} \dfrac{d}{dz}V = 0\;.
\label{eq:zone1}
\end{equation}
We have assumed that $\kappa_\mathrm{\parallel}=1/\sigma_\parallel$ is independent of $z$, which is a reasonable lowest-order approximation in the extra-planar region explored in our simulations ($\vert z \vert < 3.5$~kpc; see Figure~5 in \citetalias{Armillotta+22}). 

We define a gas acceleration scale height $\Ha$ via 
\begin{equation}\label{eq:Hadef}
d_z V \equiv \frac{V}{\Ha}\, .  
\end{equation}
Combining with $d_z P_\mathrm{c} \equiv - P_\mathrm{c}/H_\mathrm{c}$, 
for $H_\mathrm{c}$ the CR pressure scale height introduced in \autoref{eq:piecewisePc}, and inserting in  \autoref{eq:zone1}, we obtain the solution for $H_\mathrm{c}$:
\begin{equation}
H_\mathrm{c} = \dfrac{3}{8} \Ha \left ( 1 + \sqrt{1 + \dfrac{16}{9} \dfrac{\kappa_\parallel}{\Ha V}} \right)\;.
\label{eq:Hc}
\end{equation}
Physically, a higher diffusion rate (large $\kappa_\parallel$) tends to smooth out the vertical CR pressure profile (larger $H_\mathrm{c}$), while rapid gas acceleration (large $V/\Ha$) tends to steepen the profile (smaller $H_\mathrm{c}$).
If CRs and thermal gas are tightly coupled ($\kappa_\parallel \ll \Ha V$), $H_\mathrm{c} \approx (3/4) \Ha$, meaning that the CR scale height is set by the velocity gradient and is independent of CR energy. In the opposite limit, if the diffusion coefficient is very large ($\kappa_\parallel \gg \Ha V$), $H_\mathrm{c} \approx 0.5 (\kappa_\parallel \Ha/V)^{1/2}$.  That is, the CR scale height is set by the 
geometric mean of the CR diffusion coefficient, which smooths out the CR pressure profile, and the gas acceleration, which steepens it with $z$. In the extra-planar region, where gas is mostly ionized, NLL is the dominant damping mechanism, and with $\kappa_\parallel \propto p^{(\gamma - 3)/2}$ (\autoref{eq:kappa_scaling}) we obtain  $H_\mathrm{c} \propto p^{(\gamma - 3)/4}$ in the diffusion-dominated limit. 

Using \autoref{eq:Hc} in \autoref{eq:veff1} and \autoref{eq:keff1}, we obtain
\begin{equation}\label{eq:veffsol}
v_\mathrm{c,eff} = V\left(\frac{5}{8} + \frac{3}{8}\sqrt{1+ \frac{16}{9}\frac{\kappa_\parallel}{\Ha V} }  \right)
\end{equation} 
and 
\begin{equation}\label{eq:keffsoln}
\kappa_\mathrm{eff} = \frac{3}{2}\Ha V\left(1 + \sqrt {1+ \frac{16}{9}\frac{\kappa_\parallel}{\Ha V} }  +  \frac{2}{3}\frac{\kappa_\parallel}{\Ha V}  \right)\, .
\end{equation}
\autoref{eq:veffsol} has limiting forms
\begin{equation}\label{eq:veffcases} 
v_\mathrm{c,eff} \approx 
 \begin{cases}
			 V &  \text{for}\  \kappa_\parallel \ll \Ha V \\
         \dfrac{1}{2}\left(\dfrac{\kappa_\parallel V}{\Ha}\right)^{1/2} & \text{for}\  \kappa_\parallel \gg \Ha V  
\end{cases}
\end{equation} 
where the large diffusion limit results in a scaling 
$v_\mathrm{c,eff} \propto p^{(\gamma - 3)/4} $
from \autoref{eq:kappa_scaling} in the NLL regime.  
For the low- and high-diffusion limits, we have $\kappa_\mathrm{eff} = 3 \Ha V$ and $\kappa_\mathrm{eff} = \kappa_\parallel $, respectively, in the extraplanar region. In the low-diffusion limit, the effective diffusion coefficient $\kappa_\mathrm{eff}$ can be understood as the rate of spatial spreading of CRs caused by spatial gradients in the advection velocity or magnetic field configuration, rather than scattering due to microscopic diffusion.

An expression for $P_\mathrm{0}$ is derived by matching the fluxes of zone~0 and zone~1 (obtained by substituting $d_z P_{\mathrm{c},j} \equiv - P_\mathrm{c}/H_\mathrm{c}$ in \autoref{eq:CRflux1D}) at $z=z_\mathrm{t}$:
\begin{equation}
P_\mathrm{0} = \dfrac{F_\mathrm{in}}{4 V +{\kappa_\parallel}/{H_\mathrm{c}}} \equiv \dfrac{F_\mathrm{in}}{4 v_\mathrm{c,eff}} \;,
\label{eq:P0}
\end{equation}
where the second equivalence comes from \autoref{eq:veff1}.

\autoref{eq:P0} and \autoref{eq:veffcases} show that if the diffusion coefficient is low ($\kappa_\parallel \ll \Ha V$), the CR spectrum measured at the midplane would be the same as the injection spectrum, $P_\mathrm{0} \propto p^{4 -\gamma_\mathrm{inj}}$, i.e $\gamma = \gamma_\mathrm{inj}$. 
In the limit of rapid diffusion ($\kappa_\parallel \gg \Ha V$), the expected spectrum of the pressure at the midplane is $P_\mathrm{0} \propto p^{4 -\gamma_\mathrm{inj} - (\gamma-3)/4}$. 
Since $P_\mathrm{0} \propto p^{4-\gamma}$ by definition (see \autoref{eq:jth-energy}),  
this implies $\gamma_\mathrm{inj} + (\gamma-3)/4 =\gamma$, so that 
\begin{equation}\label{eq:gamma}
\gamma = \frac{4}{3} \gamma_\mathrm{inj} -1 . 
\end{equation}
Thus, for CR energies that satisfy the high diffusion limit $\kappa_\parallel \gg \Ha V$ in the extraplanar region, $\gamma = 4.73$ is expected for the measured slope of the midplane pressure when the input slope is $\gamma_\mathrm{inj} = 4.3$. This predicted value of $\gamma$ is consistent with the spectral slope of CRs with $E_\mathrm{k} > 10$~GeV detected from the Earth ($\gamma \approx 4.7$). 
In the high diffusion limit, 
$H_\mathrm{c}  \approx H_a v_\mathrm{c,eff}/V$,
so that using the second case in \autoref{eq:veffcases}, the first case in \autoref{eq:kappa_scaling}, and \autoref{eq:gamma}  we obtain  
$v_\mathrm{c,eff}\propto p^{(\gamma-3)/4} \propto p^{\gamma_\mathrm{inj}/3 - 1}$. With $\gamma_\mathrm{inj}=4.3$, this corresponds to $H_\mathrm{c} \propto v_\mathrm{c,eff} \propto p^{0.43}$. 

We can also apply our two-zone model to derive an analytic prediction for CR grammage.
The grammage is a measure of the total column of gas traversed by CRs as they propagate through the ISM, defined for an individual CR particle as $X = \int \rho v_\mathrm{p} dt$. Since most of the mass resides within the warm/cold ISM, CRs accumulate the majority of their grammage crossing the disk midplane (which occurs multiple times prior to escape). Thus, we can approximate the grammage as $X \approx v_\mathrm{p} \rho_\mathrm{ISM}   t_\mathrm{esc} $, where $\rho_\mathrm{ISM}$ is the mean gas density in the ISM, and $  t_\mathrm{esc}$ is the mean escape time of CRs from the galactic disk. The latter can be expressed as $ t_\mathrm{esc} = E_\mathrm{midplane}/\dot{E}_\mathrm{out} \approx  3 P_\mathrm{0} (2 H_\mathrm{gas} A)/\dot{E}_\mathrm{out}$, where $H_\mathrm{gas}$ is the disk scale height, $A$ is the disk area, $E_\mathrm{midplane}=6 P_\mathrm{0} H_\mathrm{gas} A$ is the total energy of CRs (of a given momentum) in the disk midplane region, and $\dot{E}_\mathrm{out} \approx 2 A F_\mathrm{c}(z_\mathrm{t}) = 2 A F_\mathrm{in}$ is the rate of CR energy leaving the disk. This results in the following expression for the grammage:
\begin{equation}
X \approx 3 \dfrac{P_0}{F_\mathrm{c}(z_\mathrm{t})} v_\mathrm{p} \rho_\mathrm{ISM} H_\mathrm{gas}  = \dfrac{3}{8}   \dfrac{v_\mathrm{p}}{v_\mathrm{c,eff}} \Sigma_\mathrm{gas}\;,
\label{eq:grammage_approx}
\end{equation}
with $\Sigma_\mathrm{gas} = 2 \rho_\mathrm{ISM} H_\mathrm{gas}$ the mean gas surface density.  From \autoref{eq:veffcases}, one can then expect grammage to scale with CR momentum at high $p$ as 
\begin{equation}\label{eq:Xscaling}
X \propto p^{(3-\gamma)/4} \propto 
 p^{1-\gamma_\mathrm{inj}/3}. 
\end{equation}
With $\gamma_\mathrm{inj}=4.3$, the predicted scaling of grammage with momentum becomes $X \propto p^{-0.43}$.

In \autoref{sec:new1dmodel}, we discuss how the solution of the analytic model would change if we relax the assumption of CR energy components evolving independently. We demonstrate that, to first order, our solution remains largely unaffected, with only a marginal variation in the normalization of the entire spectrum.

\subsection{Comparison to the simulation}
\label{sec:comparison}

We now compare the simulation outcomes to the predictions of the analytic model. For this comparison, we use the data from the post-processing simulations rather than those from the MHD relaxation simulations because the later do not include SN feedback, which leads to a drop in gas velocities. Using post-processing profiles for CRs is justified by the fact that the mean vertical profiles of CR pressure do not vary significantly once MHD is turned on (see \autoref{fig:spectrum}).

\begin{figure*}
\centering
\includegraphics[width=\textwidth]{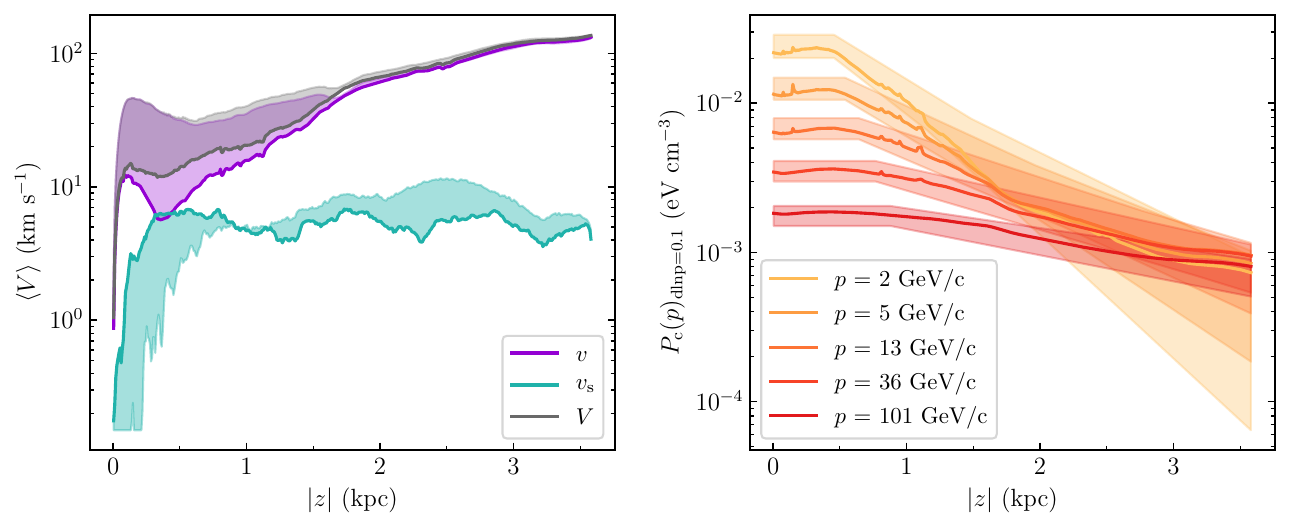}
\caption{Vertical profiles of MHD velocities (left) and CR pressure in different spectral bins (right). The left panel displays the pressure-weighted mean vertical components of the advection velocity $v_\mathrm{z}$ (violet), CR streaming velocity $v_\mathrm{s,z}$ (cyan), and MHD velocity $V_\mathrm{z} \equiv v_\mathrm{z} + v_\mathrm{s,z}$ (gray) as a function of $z$. 
The solid lines indicate the mean profiles for CRs with momentum $p=2$~GeV/c, while the shaded areas cover velocity variations at different momenta. 
The combined MHD velocity is used to evaluate the parameter $\overline{d_zV}$ employed in the 1D model.
The right panel compares the time- and horizontally-averaged profiles of CR pressure 
$P_\mathrm{c}$ obtained in the simulation (solid lines) to the CR pressure profiles predicted by the 1D model (shaded areas), with different colors corresponding to different CR momentum bins. The shaded areas represent variations in the model parameters $\overline{V}$ ($10-15 \,\kms$) and $\overline{d_zV}$ ($15-27 \,\kms$~kpc$^{-1}$).}
\label{fig:vz&Pc}
\end{figure*} 

The values of $\kappa_\parallel$ and $V$ applied in the model (hereafter $\overline{\kappa}_\parallel$ and $\overline{V}$) are evaluated in the regime where CR transport is most limited, which occurs at temperature $T\gtrsim 10^4$~K, where gas transitions from being mostly neutral to mostly ionized. In this regime, the diffusion coefficients decrease sharply, leading CRs to couple with the gas. 
In practice, for each CR energy bin, we set $\overline{\kappa}_\parallel$ to the minimum value of its median diffusion coefficient profile, occurring at $10^4 < T < 5 \times 10^4$~K, as shown in the left panel of \autoref{fig:sigma_vel}. The dependence of $\overline{\kappa}_\parallel$ on $p$ is well described by a power law: $\overline{\kappa}_\parallel \approx 7.2 \times 10^{27} [p/(3\, \mathrm{Gev/c})]^{0.9}$~cm$^2$~s$^{-1}$. This is slightly steeper than what expected if $\kappa$ only depends on $p$ through $n_1$, which would lead to 
$\kappa_\parallel \propto p^{0.8}$ for $\gamma = 4.63$ (the best-fit slope of our average CR distribution function, see right panel of \autoref{fig:spectrum}). The stronger observed dependence arises because the local CR scale height along the magnetic field direction ${P_\mathrm{c}}/{\vert \mathbf{\hat{B}} \cdot \nabla  P_\mathrm{c}\vert}$, which appears in the scattering coefficient equation (\autoref{eq:sigmaNLL}), also has a slight dependence on $p$; specifically, it increases with $p$ as the CR distribution becomes smoother. The gas MHD vertical velocity $\overline{V}$ (combining advection and Alfv\'en streaming speeds) is evaluated at the same temperatures where the diffusion coefficients reach their minimum values: $10 < V < 15\,\kms$ for $10^4 < T < 5 \times 10^4$~K, as shown in \autoref{fig:sigma_vel}. 

The gas acceleration scale $\Ha$ is defined as $\overline{V}/\overline{d_z V}$, with $\overline{d_z V}$ the mean gas acceleration. The latter is computed by performing a linear fit to the CR pressure-weighted average vertical profile of the $z$-component of the MHD velocity, shown in gray in the left panel of \autoref{fig:vz&Pc}. We find that $\overline{d_z V}$ varies between 15 and 27 $\kms$~kpc$^{-1}$ depending on the CR momentum and the range of $z$ used for the fit. This results in $\Ha$ varying between $\sim 0.4$ and 1~kpc. In the left panel of \autoref{fig:vz&Pc}, we also display the mean vertical profiles of the $z$-components of the gas advection velocity and CR streaming velocity. We note that the CR pressure-weighted advection and streaming velocities are comparable for GeV CRs in the region where they transition from zone~0 to zone~1 ($\approx 500$~pc; see also left panel of \autoref{fig:spectrum}). However, as CR momentum increases, the advection velocity becomes significantly higher than the streaming velocity. This difference arises because lower-energy CRs are primarily concentrated in the warm/cold gas phase, which has relatively low advection velocities, while higher-energy CRs are more evenly distributed across different gas phases. In comparing the model to the simulations, we allow for a range of $V$ and $\Ha$ to accommodate these measured variations. 

\begin{figure*}
\centering
\includegraphics[width=\textwidth]{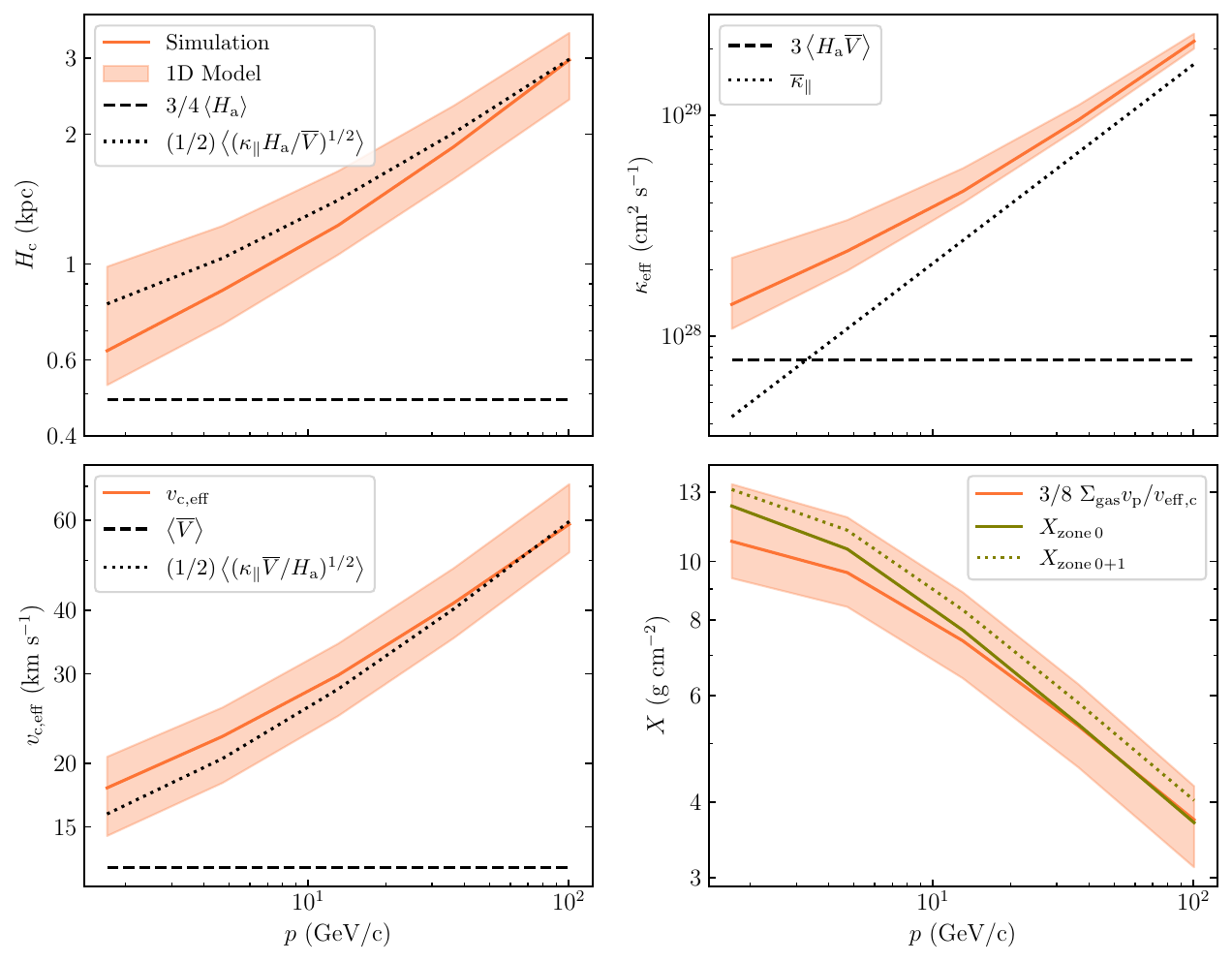}
\caption{Comparison of key CR transport properties as predicted by the simulation (solid lines) versus the 1D model (shaded areas). The top left, top right, bottom left, and bottom right panels show, respectively, the CR scale height $H_\mathrm{c}$, effective diffusion coefficient $\kappa_\mathrm{eff}$, effective velocity $v_\mathrm{c,eff}$, and grammage $X$, each as a function of momentum $p$. The shaded areas cover variations in the model parameters $\overline{V}$ ($10-15 \,\kms$) and $\overline{d_zV}$ ($15-27\, \kms$~kpc$^{-1}$). 
Horizontal dashed black lines indicate asymptotic limits that apply for low diffusion, and diagonal dotted black lines indicate asymptotic limits that apply for high diffusion, as marked in panel keys.  
In the bottom right panel, the orange line and the shaded area are computed using the approximate formula for the grammage (\autoref{eq:grammage_approx}). The solid and dotted green lines represent the grammage directly derived from the simulation data (Equation~27 of \citetalias{Armillotta+21}) evaluated in the disk region ($\vert z \vert < 500$~pc) or across the entire simulation box, respectively.}
\label{fig:1dmodel}
\end{figure*}

The agreement between the simulation outcomes and the prediction of the 1D model is remarkably good, as we demonstrate in the right panel of \autoref{fig:vz&Pc} and in \autoref{fig:1dmodel}. The right panel of \autoref{fig:vz&Pc} shows the mean vertical profiles of CR pressure within spectral bins with equal size, $d\mathrm{ln}p = 0.1$, for different CR momenta $p$. The simulated profiles are obtained by taking horizontal averages across the analyzed TIGRESS snapshots. The analytic profiles -- represented with shaded areas to cover variations in the parameters $\overline{V}$ and $\overline{d_z V}$ -- individually take the form in \autoref{eq:piecewisePc}.
Based on our measured $\overline\kappa_\parallel$, $\overline V$, and $\overline{d_zV}$ values in each energy bin, we compute $H_\mathrm{c}$ from \autoref{eq:Hc} and $P_0$ from \autoref{eq:P0}. The transition point $z_\mathrm{t}$ represents the mean height at which gas transitions from neutral to ionized. As the ISM is highly inhomogeneous and time-variable, this occurs over a range of values. We therefore simply fit \autoref{eq:piecewisePc} to the simulated CR pressure profiles to obtain $z_\mathrm{t}$, with a range $500-800$~pc. 

\autoref{fig:1dmodel} displays key CR transport properties as a function of $p$. The top left panel compares the scale heights obtained by fitting \autoref{eq:piecewisePc} to the simulated pressure profiles in each energy bin with the model prediction in \autoref{eq:Hc}. The top right and bottom left panels show, respectively, the comparison between the effective diffusion coefficients $\kappa_\mathrm{eff}$ (\autoref{eq:keff1}) and effective CR velocities $v_\mathrm{c,eff}$ (\autoref{eq:veff1}) 
obtained using the fitted CR scale heights for $H_\mathrm{c}$ versus those derived using 
\autoref{eq:Hc} for $H_\mathrm{c}$ (i.e. \autoref{eq:veffsol} and \autoref{eq:keffsoln}).
In the panels for $H_\mathrm{c}$, 
$\kappa_\mathrm{eff}$, $v_\mathrm{c,eff}$, we also plot with dashed lines the mean values of $(3/4) \Ha$, $3\Ha \overline{V}$, and $\overline{V}$, which represent the respective expected limiting values for small diffusion coefficient (as would apply for $p$ small enough that $\kappa_\parallel \propto p^{0.9} \ll \Ha V$). In these three panels, we also plot with dotted lines the limiting forms $H_\mathrm{c}=0.5(\kappa_\parallel \Ha/ \overline{V})^{1/2}$, 
$\kappa_\mathrm{eff} = \kappa_\parallel$, and 
$v_\mathrm{c,eff}= 0.5(\kappa_\parallel \overline{V}/ \Ha )^{1/2}$ 
(as would apply for $p$ large enough that  $\kappa_\parallel \propto p^{0.9} \gg \Ha V$). The comparison indicates that, as $p$ decreases, the profiles of $H_\mathrm{c}$, 
$\kappa_\mathrm{eff}$, and $v_\mathrm{c,eff}$ tend towards their advection-dominated limits, while, at high $p$, where diffusion is dominant, they approach a power-law behavior. 

Finally, the bottom right panel of \autoref{fig:1dmodel} displays the CR grammage. Here, the orange line and shaded area represent the grammage calculated using the approximate formula given by \autoref{eq:grammage_approx}, while the green lines show the grammage computed directly from the simulation data using the formula derived in \citetalias{Armillotta+21}: $X = \mu_\mathrm{H} m_\mathrm{p} v_\mathrm{p} \int d^3 x\, n_\mathrm{H} e_\mathrm{c} / \dot{E}_\mathrm{in}$,
where $\mu_\mathrm{H} = 1.4$ is the mean molecular weight per hydrogen atom. The green solid and dotted lines correspond to the grammage evaluated in the disk region ($\vert z \vert < 500$~pc) and in the total simulation box, respectively. The comparison between these two lines clearly shows that most of the grammage is accumulated within the galactic disk, in line with the assumption underlying \autoref{eq:grammage_approx}. This explains the excellent agreement between the solid green and orange lines at high CR momenta. At low momenta, however, \autoref{eq:grammage_approx} slightly underestimates the grammage. This discrepancy arises from the implicit assumption behind the 1D model that the CR energy density is constant in the $xy$-direction. While this holds for high-momentum CRs, whose distribution is uniform across gas phases, low-momentum CRs preferentially reside in the warm/cold gas. As a result, the actual pressure of low-momentum CRs in the warm/cold gas, where most of the grammage is accumulated, is slightly higher than the average CR pressure across all phases ($P_\mathrm{0}$). This results in a lower predicted grammage in the model.

We note that the simulations presented in this work -- both the postprocessing and the MHD relaxation simulations -- do not include SN and CR feedback in a fully self-consistent manner. Therefore, in  
fully self-consistent CR-MHD simulations the values of the three key parameters that enter our model and determine the vertical CR distribution -- $V$, $d_{z}V$ ($\equiv V/H_a$), and $\kappa_\parallel$ -- might differ 
from those estimated here. To asses the potential impact of this difference, we use preliminary results from a new TIGRESS simulation of the solar neighborhood environment (C.-G.~Kim et al., in prep.), which includes SN and CR feedback self-consistently, although without resolving the CR energy spectrum. In agreement with other studies, this simulation shows that CRs contribute to accelerating warm gas into the extra-planar region. Compared to the mean values adopted in this paper, the new TIGRESS simulation indicates modest increases in both $V$ and $\kappa_\parallel$($E_\mathrm{k}=1$~GeV), and a decrease in $d_{z}V$ by a factor of a few. Taken together, these differences leave the effective CR transport velocity and CR pressure nearly unchanged,
while increasing the scale height (primarily due to the larger leading coefficient in \autoref{eq:Hc_app}). 
This suggests that results from future CR-MHD simulations for the midplane CR spectra and grammage will remain consistent with our present results, and with observations in the solar system, but extraplanar CR scale heights may be larger than those estimated in the present study. Future postprocessing of the new TIGRESS simulation using the spectrally resolved CR scheme developed in this study will enable more robust estimates of the CR spectral slope and energy-dependent scale height in a more realistic environment that includes CR feedback.

\section{Discussion}\label{sec:discuss}
\subsection{Connection to phenomenological models}
\label{sec:phen_models}

Numerous studies, both analytic and numerical, have been undertaken to constrain the properties of CR transport using direct measurements of local CR primary spectra and secondary-to-primary CR ratios \citep[see reviews by][]{Strong+07, Amato&Blasi18}. A key strength of these phenomenological models is their detailed treatment of collisional processes and the resulting production of radiation and secondary particles, which is crucial for reproducing the spectra of different CR species. However, these models typically rely on simplified prescriptions for the underlying gas and magnetic field distribution, as well as the CR propagation itself. For example, they assume a spatially constant CR diffusion coefficient ${\cal D}$ that depends only on particle momentum, typically following a broken-power law functional form. Recent constraints on the slope of ${\cal D}$ with $p$ 
range between 0.4 and 0.6 for the momentum range considered in this work \citep[e.g.][]{Boschini+20, Evoli+20, Silver+24}. In the diffusion-dominated regime, which is generally found to hold for $p\gtrsim 10$~GeV/c, the grammage $X \propto {\cal D}^{-1}$.

The observationally constrained $p$-dependence of $\cal D$ is weaker than that predicted by our simulation and employed in our 1D model ($\overline{\kappa}_\parallel \propto p^{0.9}$; see \autoref{sec:comparison}). However, it is crucial to note that the mathematical form of our model differs from that of phenomenological models. One key difference is that most phenomenological models assume that CRs freely escape beyond a certain distance from the disk, and this escape distance is momentum-independent. In contrast, our model does not impose free-escape boundary conditions and instead shows that the vertical CR distribution is sensitive to $p$ (with scale height $H_\mathrm{c} \propto p^{0.43}$ at large $p$). 
Because the length scale is independent of $p$ in phenomenological models, a more meaningful comparison is between the momentum dependence of $\cal D$ and the momentum dependence of  $\kappa_\mathrm{eff}/H_\mathrm{c} = 4v_\mathrm{c,eff}$ in our model, which follows $v_\mathrm{c,eff} 
\propto \kappa_\parallel^{1/2} \propto p^{(\gamma-3)/4}\propto p^{\gamma_\mathrm{inj}/3 -1 } \propto p^{0.43}$ at high $p$. Our model results therefore align closely with the estimates from phenomenological models, when properly compared. 
Equivalently, from \autoref{eq:grammage_approx} we find grammage $X\propto v_\mathrm{c,eff}^{-1} $, with measured slope $-0.33$ in our simulations, and an asymptotic prediction for the scaling of $X\propto p^{-0.43}$ from \autoref{eq:Xscaling}. 

Another distinction between our model and phenomenological models is that our model does not include collisional losses, as they are negligible for CR protons. If collisional losses were significant, \autoref{eq:CRenergy1D} for zone~0 would become  $dF_\mathrm{c}/dz = Q_\mathrm{SN} - \Lambda_\mathrm{coll} n_\mathrm{H} e_\mathrm{c}$, which implies $F_\mathrm{c}(z_\mathrm{t}) = F_\mathrm{in} - 3 P_\mathrm{c} \Lambda_\mathrm{coll} \Sigma_\mathrm{gas}/(2 \mu_\mathrm{H} m_\mathrm{p})$. By matching the fluxes of zone~0 and zone~1 (the flux of zone~1 remains unchanged as collisional losses are always negligible for protons in the low-density extra-planar region), $P_\mathrm{0}$ would become:
\begin{equation}
P_\mathrm{0} = \dfrac{F_\mathrm{in}}{4 v_\mathrm{c,eff} + \dfrac{3 \Lambda_\mathrm{coll} \Sigma_\mathrm{gas}}{2 \mu_\mathrm{H} m_\mathrm{p}}} \approx \dfrac{F_\mathrm{in}}{4 v_\mathrm{c,eff} \left (1+ \dfrac{\Lambda_\mathrm{coll} X}{\mu_\mathrm{H} m_\mathrm{p} v_\mathrm{p}} \right )} \;,
\label{P0_w_losses}    
\end{equation}
where we have used \autoref{eq:grammage_approx} for the grammage $X$, and where $F_\mathrm{in}=\dot E_\mathrm{in}/(2A)$ is the CR flux injected in a given momentum bin. With \autoref{eq:veffsol} for $v_\mathrm{c,eff}$, our model provides a simple prediction for the energy dependence of observed CR protons. The quantities $\Ha$ and $V$ can be predicted from MHD simulations, but this form may also be useful more generally in phenomenological modeling, to test best-fit values of $\Ha$ and $V$.  

\citet{Kempski+22} argue that the predictions of phenomenological models are incompatible with the self-confinement scenario. Their argument is based on a one-zone model for vertical CR transport, derived from the Fokker-Planck equation for the CR distribution function; this is essentially equivalent to solving the system described by \autoref{eq:CRenergy1D} and \autoref{eq:CRflux1D} in a single zone. They argue that, if linear damping mechanisms, such as IN damping, dominate, CR transport transitions sharply from the advection/streaming regime to the speed-of-light propagation regime. Thus, linear damping by itself would not produce energy-dependent transport. In contrast, if nonlinear damping mechanisms, such as NLL damping, dominate, CR transport becomes energy-dependent, but for the \citet{Kempski+22} one-zone model the dependence is stronger than what is found in phenomenological models. 

Our two-zone model rectifies the apparent failure of the self-confinement scenario in reproducing the observations. In the IN damping regime that applies within the mostly neutral midplane layer, the CR pressure is given by \autoref{eq:P0}, which is similar in form to the equation derived by \citet{Kempski+22} for linear damping (cf.~their Equation~2). However, the diffusion coefficient in \autoref{eq:P0} is not the midplane value, but instead the value typical of the highly ionized galactic-fountain gas, where NLL damping dominates (it is the low-efficiency transport in this region that controls the propagation of CRs out of the midplane layer; see also semi-analytical model by \citealt{Chernyshov+22}). Moreover, based on our simulations, the global CR scale height in this extraplanar region is itself momentum dependent, varying as $H_\mathrm{c} \propto \kappa_\parallel^{1/2}$ at large momentum in our two-zone model (see \autoref{eq:Hc}). 
These differences ensure that the effective velocity controlling CR flows away from the midplane  is energy-dependent in our model, with $v_\mathrm{c,eff} \propto \kappa_\parallel^{1/2}$ at large momentum. In the NLL regime of the ionized fountain gas, our \autoref{eq:zone1} differs from Equations~12-14 in \citet{Kempski+22} in that we do not include a term for CR injection, as there are no sources in the extraplanar region, but we do include terms associated with acceleration of the flow and adiabatic losses. It is worth noting also that the local CR gradient scale along the magnetic field, which enters in defining $\kappa_\parallel$ (based on \autoref{eq:sigmaNLL}), differs from the global vertical CR scale height and the former is weakly dependent on $p$. In the one-zone model of \citet{Kempski+22} there is no distinction between local and global CR gradients.

\subsection{Comparison to other CR-MHD simulations}

In recent years, several studies have focused on modeling the transport of spectrally resolved CRs in MHD simulations of galaxies or portions of ISM (see \autoref{sec:introduction}). Similarly to our work, a few of these studies have specifically investigated the transport of CR protons in simulations of galaxies with conditions representative of the Milky Way \citep[e.g.,][]{Hopkins+22a, Hopkins+22, Girichidis+24}. The treatment of CR transport in these studies differs from ours in that they cover a broader spectral range, extending to non-relativistic energies, and provide a more accurate modeling of CR transport in momentum space, while we evolve CRs in each momentum bin independently. Furthermore, these studies perform fully self-consistent simulations with time-dependent MHD and CR physics, while we conduct postprocessing simulations followed by a short MHD relaxation step.

The novelty of our study, compared to previous works, lies in the use of a physically motivated prescription for variable CR scattering in high-resolution simulations of the dynamic, multiphase ISM. This approach allows us to explore the predictions of the self-confinement theory for CR transport and compare the results against observational data. In contrast, earlier studies employ a spatially constant, energy-dependent scattering coefficient, which is either based on empirical estimates from phenomenological models, or calibrated to match the simulated spectra with those observed in the local ISM. The only other study to test the self-confinement theory using a variable scattering model is that of \citet{Hopkins+22}, which employs cosmological zoom-in FIRE simulations \citep{Hopkins+18}. However, in contrast to our findings, they find that the predictions of the standard self-confinement scenario are inconsistent with the observations. Specifically, they conclude that the self-confinement theory fails to reproduce both the normalization and the slope of CR spectra. Using an analytic argument similar to that of \citet{Kempski+22}, they attribute the lack of agreement between their simulations and the observations to a ``solution collapse'' problem in the self-confinement theory, where only two steady-state solutions are possible: the system either collapses to the infinite scattering limit, where CRs can only propagate at the Alfv\`{e}n speed, or to the free-streaming limit with no scattering, where CRs propagate at the speed of light.

In \citetalias{Armillotta+21} and \citetalias{Armillotta+24}, we discussed some differences between our simulations and the FIRE simulations with CRs, which could help explain the failure of the self-confinement model in \citet{Hopkins+22}. These differences include orders of magnitude coarser mass resolution of the hot gas, which prevents the FIRE simulations from properly resolving the hot phase of the ISM. As highlighted at the end of \autoref{sec:phen_models}, a two-zone transport model, in contrast to a one-zone model, agrees well with the numerical results, and provides a physical understanding of how self-confinement works in a multi-phase disk.   
In order to accurately model CR transport in numerical MHD simulations, a 
realistic representation of the dynamical and thermal properties of the ISM is necessary, which  requires high spatial resolution.

\section{Summary}\label{sec:summary}

This study examines the transport of CR protons with kinetic energy ranging from 1 to 100 GeV within the dynamic, multiphase ISM. Building on our previous studies (\citetalias{Armillotta+21, Armillotta+22, Armillotta+24}), which considered single-energy GeV CRs, we now compute the transport of spectrally resolved CRs in a TIGRESS simulation for an environment similar to the solar neighborhood \citep{Kim&Ostriker17, Kim+20}. 
Our simulations follow spatial transport of 
5 distinct CR proton components independently, each representing a specific range of momenta and treated as a relativistic fluid. The energy density and flux of each component is evolved using the scheme developed in our previous works. A key feature of our CR transport model is the space- and momentum-dependent scattering coefficient, set by the balance between streaming instability growth and NLL/IN damping. The CR spectral distribution is initialized near source particles, assuming that $10\%$ of SN energy goes into acceleration of CR protons with $p \geq 1$~GeV/c, and following an injection spectrum $ f_\mathrm{inj}(p)\propto p^{-4.3}$. 

Our simulations show that as CR protons propagate away from their sources through the ISM, their spectrum steepens due to momentum-dependent diffusion. The average spectrum in the disk region aligns well with the CR proton spectrum measured in the solar system. The simulated spectrum closely follows a power law $f(p) \propto p^{-\gamma}$ with a slope of $\gamma \approx 4.6$, which is in excellent agreement with the observed slope of $4.7$ for CR protons with kinetic energies $E_\mathrm{k} > 10$~GeV \citep[e.g.,][]{Aguilar+14, Aguilar+15}. 

Consistent with our previous findings, the spatial distribution of CRs is nearly homogeneous in the neutral warm/cold gas that dominates the disk midplane region, since the scattering rate is quite low where ion-neutral collisions strongly damp waves. In the extra-planar region, where the gas has much lower density and higher ionization, wave damping is much reduced, and the CR scattering is enhanced. In the extraplanar region the density of all CR components drops exponentially with $\vert z \vert$. The  structure of a highly diffusive midplane sandwiched between lower diffusion extraplanar regions holds across all momentum values, although the extraplanar CR gradient is less pronounced at higher momenta due to increased diffusion rates.

We introduce a novel two-zone analytic model for vertical CR propagation, which allows for CRs to be transported by the magnetized outflow and also to diffuse relative to the gas. Crucially, the values of the diffusion coefficient $\kappa_\parallel$ differ in the neutral midplane gas and ionized extraplanar gas, and depend on momentum following the self-confinement prescription.  
Our simulation findings for vertical pressure profiles are in excellent agreement with the prediction of the analytic model. In all momentum bins, profiles are flat in the disk midplane region (zone~0), and exponential in the extra-planar region (zone~1). The pressure in zone~0 is $P_0 \approx F_\mathrm{in}/(4 v_\mathrm{c,eff})$ for $F_\mathrm{in}$ the input flux and $v_\mathrm{c,eff}$ the effective CR transport velocity in zone~1 (\autoref{eq:veff}), which depends on the the sum of the mean gas velocity and Alfv\'en speed $V$, acceleration scale $\Ha$, and CR diffusion coefficient in zone~1 
(see \autoref{eq:veffsol}). 
The scale height $H_\mathrm{c}$ of the pressure profile in the extra-planar region also depends on these three factors (\autoref{eq:Hc}). Specifically, at low momenta, CR transport is primarily controlled by gas advection ($v_\mathrm{c,eff} \rightarrow V$, $H_\mathrm{c} \rightarrow (3/4) \Ha$), while at high momenta, transport  
depends on both advection and diffusion ($v_\mathrm{c,eff} \rightarrow 0.5 ( \kappa_\parallel V/ \Ha )^{1/2}\propto p^{(\gamma-3)/4}$, $H_\mathrm{c} \rightarrow 0.5 (\kappa_\parallel \Ha/V)^{1/2} \propto p^{(\gamma-3)/4}$).
Using our analytic model, we show that the spectral slope is expected to be related to the injection slope by $\gamma = (4/3)\gamma_\mathrm{inj} -1$, in good agreement with the results of our simulations. This implies that the steepening $\Delta \gamma=\gamma-\gamma_\mathrm{inj} = (\gamma_\mathrm{inj} - 3 )/3$ will be close to 1/3 for $\gamma_\mathrm{inj}\simeq4$.

Taken together, the excellent agreement between our simulations, our analytic model, and observations strongly supports a new conception of CR transport. In this picture, CRs are confined within the mostly-neutral disk midplane by the surrounding higher-ionization gas, which is able to support waves that resonate with, and therefore scatter, CRs.  At each CR momentum, the effective velocity of the flow out of the midplane region is controlled both by the (momentum-dependent) scattering rate in the diffuse ionized gas that sandwiches it, and the outward acceleration of that gas.  
We conclude that both the multiphase character of the ISM, and its large-scale dynamics, must be taken into account for a physically realistic treatment of CR transport. 

The present work demonstrates that the combination of the TIGRESS star-forming ISM framework and our CR implementation is able to properly capture the steepening between input and ambient spectra, and to obtain realistic energy densities and grammage for CR protons the $1-100$ GeV regime. In a separate work (Linzer et al 2025, accepted), we present results for energy-dependent CR electron transport in the same energy range. There, we show that electron spectra steepen more than proton spectra due to additional losses, with (energy-dependent) electron spectral slopes in good agreement with observed constraints. In combination, these studies provide important validation of the CR transport scheme we have implemented. More generally, our work provides support for the approach in which scattering coefficients (in the $1-100$ GeV regime) in CR-MHD simulations are computed locally based on a balance of wave excitation and damping. This approach is straightforward to implement, and limits the need for free parameters in numerical simulations that investigate dynamical  consequences of CRs in the ISM, galactic winds, and the circumgalactic medium.  When adopting this approach, the fractional ionization must also be computed locally (based on ionizing sources including the CRs themselves), since wave damping depends strongly on the ionization fraction.     

\section*{Acknowledgements}
We are grateful to the referee for a helpful and constructive report. This work was performed in part at the Aspen Center for Physics, which is funded by the National Science Foundation (NSF) under grant PHY-2210452. Support for this work was provided by grant 510940 from the Simons Foundation to ECO and grant AST-2407119 from the NSF to LA and ECO. LA was supported in part by the INAF Astrophysical fellowship initiative.

\bibliography{bib}{}
\bibliographystyle{aasjournal}

\appendix

\section{Additional details of the cosmic-ray transport scheme}

\subsection{Calculation of $n_\mathrm{1}$ in the scattering coefficient formula}
\label{Appendix_n1}

For each CR fluid component (index $j$), the quantity $n_{1,j}$ depends on the resonant momentum $p_{1,j}$, which is equal to the momentum $p_{j}$ associated with that fluid component, and on the local CR distribution function $f(p)$ (see \autoref{eq:n1}). For a generic distribution function $f = C p^{-\gamma}$, where $C$ is the normalization factor and $\gamma$ is the slope of the distribution function, $n_{1,j}$ can be expressed as
\begin{equation}
n_{1,j} = 4 \pi C \dfrac { p_{j}^{3-\gamma}}{\gamma-2} \;,
\label{eq:n1_appendix}
\end{equation}
provided $\gamma>2$, which is always the case in our simulations. Assuming that, in \autoref{eq:jth-energy}, $p$ and $E$ are constant within the spectral range $d\mathrm{ln}p$, we approximate $C \approx e_{\mathrm{c},j} / (4 \pi p^{4-\gamma} c d\mathrm{ln}p)$. Thus, in each cell, $n_{1,j}$ is computed as
\begin{equation}
n_{1,j} = \dfrac {e_{\mathrm{c},j}}{E_\mathrm{j} (\gamma-2) d\mathrm{ln}p} \;.
\label{eq:n1_code}
\end{equation}
For the purpose of calculating the scattering coefficient, we do not compute $\gamma$ locally in each cell; instead, we assume $\gamma = 4.7$, consistent with the slope of the distribution function of CR protons detected from the Earth \citep[e.g.,][]{Aguilar+14, Aguilar+15}, which is also within 2\% of the mean value $\gamma\approx 4.6$ found in our simulations.

\subsection{Collisional Losses}
\label{Appendix_loss}
In \tworef{eq:CRenergy}{eq:CRflux}, $\Lambda_{\mathrm{coll},j} n_\mathrm{H} e_{\mathrm{c},j}$ and $\Lambda_{\mathrm{coll},j} n_\mathrm{H} \mathbf{F}_{\mathrm{c},j}/v_{\mathrm{p},j}^2$ represent, respectively, the rates of CR energy density and CR energy flux  lost via collisional interactions with the ambient gas. The energy-loss rate coefficient $\Lambda_\mathrm{coll}$ is a function of the CR energy $E$, and $\Lambda_{\mathrm{coll}, j}$ represents the value of  $\Lambda_\mathrm{coll}$ at the energy value $E_{j}$ associated to the $j$-th CR fluid.  $\Lambda_\mathrm{coll}$ is defined as  $\Lambda_\mathrm{coll}(E) = v_\mathrm{p} L(E)/E$, where $L(E)$ is the energy-loss function, defined as the product of the energy lost per collision event and the cross section of the collisional interaction.

The loss function $L(E)$ can depend on one or more collisional processes. For CR protons with $E_\mathrm{k} \gg 1$ GeV, the dominant loss mechanism is pion production caused by elastic collisions with the surrounding atoms. At $E_\mathrm{k} \sim 1$ GeV, the main loss mechanism of CRs in mostly neutral gas is ionization of atomic and molecular hydrogen, while losses due to Coulomb interactions prevail in ionized gas. In our simulations, we account for all three loss mechanisms. The pion production loss function $L_\mathrm{pion}$ at $E_{\mathrm{k}} \geq 10$~GeV is derived from \citet{Krakau+15} and adjusted by a factor 1.18 to account for collisions with particles heavier than hydrogen \citep[see][]{Padovani+20}. Below 10~GeV, we extrapolate $L_\mathrm{pion}$ using a power-law function. The corresponding energy-loss rate coefficient is:
\begin{equation}
    \Lambda_\mathrm{coll, pion} =  \begin{cases}
        4.54 \times 10^{-16} \left(\dfrac{E}{\mathrm{GeV}}\right)^{0.28} \left(\dfrac{E}{\mathrm{GeV}} + 200 \right)^{-0.2}\textrm{ cm}^{3}\textrm{ s}^{-1} \;\; & \text{if } E_{\rm{k}} \geq 10 \textrm{ GeV} \\
        3.33 \times 10^{-15} \left (
        \dfrac{\mathrm{GeV}}{E} \right) \left(\dfrac{E_{\rm{k}}}{10\textrm{ GeV}}\right)^{1.28}\textrm{ cm}^{3}\textrm{ s}^{-1} & \text{if } E_{\rm{k}} < 10 \textrm{ GeV} \;.
    \end{cases}
\label{eq:Lpion}    
\end{equation}

The ionization loss function $L_\mathrm{ion}$ is computed using the
Bethe-Bloch formula \citep[e.g.][]{Draine11}, 
\begin{equation}
    \Lambda_\mathrm{coll, ion} =  1.1\dfrac{x_\mathrm{n}}{E} \frac{4 \pi e^4}{m_\mathrm{e} v_\mathrm{p}}\left[\textrm{ln}\left(\dfrac{2 m_\mathrm{e} v_\mathrm{p}^2}{E_\mathrm{ion}(1-\beta^2)}\right) - \beta^2\right] \textrm{ cm}^{3}\textrm{ s}^{-1}\;,
\end{equation} 
with $E_\mathrm{ion}$ the hydrogen ionization energy, $m_\mathrm{e}$ the electron mass, $\beta = v_\mathrm{p}/c$, and $x_\mathrm{n}$ the fraction of neutrals, defined as the number density of neutrals divided by the total hydrogen number density (see Section 2.2.1 in \citetalias{Armillotta+21}), and where we apply a factor 1.1 to account for composition \citep{Padovani+20}.

Finally, the energy-loss rate coefficient due to Coulomb interactions is derived from \citet{Gould75} \citep[see also][]{Werhahn+21}:
\begin{equation}
    \Lambda_\mathrm{coll, coul} =\dfrac{x_\mathrm{e}}{E}\frac{3 \sigma_\mathrm{T} m_\mathrm{e} c^3}{2\beta} \left[\textrm{ln}\left(\dfrac{2 \gamma m_\mathrm{e} c^2 \beta^2}{\hbar \omega}\right) - \frac{\beta^2}{2}\right] \textrm{ cm}^{3}\textrm{ s}^{-1}\;,
\end{equation} 
where $\sigma_\mathrm{T} = 6.65 \times 10^{-25}$ cm$^2$ is the Thomson cross section, $\hbar$ is the Planck constant, $\omega_\mathrm{pl} \equiv \sqrt{{4 \pi e^2 x_\mathrm{e}n_\mathrm{H}}/{m_\mathrm{e}}}$ is the plasma frequency, with $x_\mathrm{e}$ the electron fraction and $n_\mathrm{H}$ the hydrogen density. The electron fraction, defined as the ratio between the number density of electrons and the total hydrogen number density is computed using Equation 24 of \citetalias{Armillotta+21} for gas at $T<2\times 10^4$~K, and the values tabulated in \citet{Sutherland&Dopita93} for gas at $T\geq 2\times 10^4$~K.

\section{Two-moment equations for the transport of interacting cosmic-ray fluids}
\label{Appendix_neweq}

In \autoref{sec:algorithm}, we note that the CR transport scheme employed in this study is approximate, as it treats each CR energy component independently (essentially based on an adaptation of the energy-integrated two-moment CR equations). A more accurate version of \tworef{eq:CRenergy}{eq:CRflux}, which accounts for CR energy transfer between fluids, can be obtained by taking moments of the Vlasov equation for the CR distribution function averaged over gyromotion -- as derived in \citet{Skilling75} -- and evaluated in the relativistic limit ($v_\mathrm{p} \approx c$). However, instead of evaluating the moments over the entire CR momentum range from $0$ to $\infty$ (which would lead to the single-bin equations of \citealt{Jiang&Oh18}), the moments are computed within the momentum boundaries of each CR bin centered at $p_{j}$.~A detailed derivation of the new equations and their implementation within \textit{Athena}++ is deferred to a future work (Armillotta \& Ostriker, in prep.). In this paper, we simply provide the final equations for the evolution of CR energy density and energy flux, with the intention of highlighting the differences relative to the current scheme. 

The equations for the $j$-th CR fluid are as follows:
\begin{equation}
\frac{\partial e_{\mathrm{c},j}}{\partial t}  + \mathbf{\nabla} \cdot \mathbf{F_\mathrm{c,\mathit{j}}} =  -(\gamma_{j}-3) (\mathbf{v} \, + \, \mathbf{v_\mathrm{s,\mathit{j}}} ) \cdot 
\tensor{\mathrm{\sigma}}_\mathrm{tot,\mathit{j}} \cdot   \left( \mathbf{F_\mathrm{c,\mathit{j}}} 
- \dfrac{\gamma_{j}}{3} \, \mathbf{v} e_{\mathrm{c},j}  \right) 
- (\gamma_j-\alpha_j-3) \,\Lambda_\mathrm{coll,\mathit{j}} n_\mathrm{H} e_{\mathrm{c},j} + Q_\mathrm{SN,\mathit{j}}
\;,
\label{eq:CRenergy_app}
\end{equation}
\begin{equation}
\frac{1}{v_\mathrm{m}^2} \frac{\partial \mathbf{F_\mathrm{c,\mathit{j}}}}{\partial t} +  \mathbf{\nabla} \cdot \tensor{\mathbf{P}}_\mathrm{c,\mathit{j}} = - \tensor{\mathrm{\sigma}}_\mathrm{tot,\mathit{j}} \cdot \left( \mathbf{F_\mathrm{c,\mathit{j}}} 
- \dfrac{\gamma_{j}}{3} \, \mathbf{v} e_{\mathrm{c},j}  \right) 
- (\gamma_j-\alpha_j-3) \, \frac{\Lambda_\mathrm{coll,\mathit{j}} n_\mathrm{H}}{v_\mathrm{p,\mathit{j}}^2} \mathbf{F}_\mathrm{c,\mathit{j}} \;.
\label{eq:CRflux_app}
\end{equation}
with
\begin{equation}
    \sigma_{\rm tot,\parallel,\mathit{j}}^{-1}= \sigma_\mathrm{\parallel,\mathit{j}}^{-1} - \gamma v_\mathrm{A,i} \frac{P_{\mathrm{c},j}}{|\hat B \cdot \nabla P_{\mathrm{c},j} |} \,,
\label{eq:sigmatotpar_app}    
\end{equation}
and $\sigma_{\rm tot,\perp,\mathit{j}} = \sigma_\mathrm{\perp,\mathit{j}}$. In deriving these equations, we assume that, within each momentum bin, both the CR distribution function and the energy-loss rate coefficient can be approximated by power laws, with $f_j \propto p^{-\gamma_j}$ and $\Lambda_\mathrm{coll,\mathit{j}} \propto p^{\alpha_j}$. For $\gamma_j = 4$ and $\alpha_j=0$, \tworef{eq:CRenergy_app}{eq:CRflux_app} reduce to \tworef{eq:CRenergy}{eq:CRflux}. In our simulations, the slope of the CR distribution function varies spatially between 4.3 and 4.6, while it changes only marginally across bins (see \autoref{sec:sims}). Additionally, in the investigated momentum range, where pion losses dominates, $ \alpha_j \approx 0.28 $ (see \autoref{Appendix_loss}). This indicates that while the equations remain very similar to those used in numerical and analytic models presented in the current paper, there would be a modest (a few tens of percent) quantitative difference in the individual source terms.

\subsection{One-dimensional model}
\label{sec:new1dmodel}

To demonstrate that employing a CR propagation scheme based on \tworef{eq:CRenergy_app}{eq:CRflux_app} would not affect the overall conclusions of the paper regarding CR spectrum dependencies, we provide the solutions of the one-dimensional model for vertical CR propagation (\autoref{sec:1Dmodel}) using \tworef{eq:CRenergy_app}{eq:CRflux_app} instead of \tworef{eq:CRenergy}{eq:CRflux}. The midplane CR pressure $P_0$ in a given spectral bin $j$ becomes:
\begin{equation}
P_\mathrm{0} = \dfrac{F_\mathrm{in}}{\gamma v_\mathrm{c,eff}} \;,
\label{eq:P0_app}
\end{equation}
with
\begin{equation}
v_\mathrm{c,eff} = V + \dfrac{\kappa_\parallel}{\gamma H_\mathrm{c}}\;,
\label{eq:vceff_app}
\end{equation}
and
\begin{equation}
H_\mathrm{c} = \dfrac{3}{2\gamma} \Ha \left ( 1 + \sqrt{1 + \dfrac{4 \gamma}{9} \dfrac{\kappa_\parallel}{\Ha V}} \right)\;.
\label{eq:Hc_app}
\end{equation}
As in \autoref{sec:1Dmodel}, we have suppressed the index $j$ for cleaner notation. By substituting \autoref{eq:Hc_app} in \autoref{eq:vceff_app}, we obtain the following expression for the effective velocity:
\begin{equation}
v_\mathrm{c,eff} = V \left(\frac{2 \gamma - 3}{2 \gamma} + \frac{3}{2 \gamma}\sqrt{1+ \frac{4 \gamma}{9}\frac{\kappa_\parallel}{\Ha V} }  \right) \;.
\label{eq:vceff_app2}
\end{equation}

These equations are qualitatively consistent with those derived in \autoref{sec:1Dmodel}. \tworef{eq:P0_app}{eq:vceff_app2} indicate that, for values of $\gamma$ close to 4, the new scheme would result in a slightly different value of $P_0$ in each momentum bin. However, in the momentum range considered in this work, the midplane proton spectrum -- both in the observations and simulations -- can be well approximated by a power law (see \autoref{sec:sims}), meaning that $\gamma$ depends only marginally on $p$, with this dependence arising from the varying relative contributions of advective and diffusive transport at different $p$. Thus, to first order, the main effect of employing this new scheme would be a change in the normalization of the entire spectrum. To second order, the effect would be a slight shift in the relative contributions of advective and diffusive transport at different $p$, leading to a marginal change of the spectral slope in the regime where both advection and diffusion are important (since $\kappa_\parallel$ depends on $p$). The slope would remain unchanged in the rapid diffusion regime (\autoref{eq:gamma}), valid at high $p$.

\end{CJK*}
\end{document}